\documentclass[%
 reprint,
 amsmath,amssymb,
 aps,
 nofootinbib,
 superscriptaddress
]{revtex4-1}

\usepackage{graphicx}% Include figure files
\usepackage{dcolumn}% Align table columns on decimal point
\usepackage{bm}% bold math
\usepackage{titlesec}
\usepackage{alphalph}
\usepackage{hyperref}
\usepackage{siunitx}
\usepackage{breqn}
\usepackage{color, soul}
\usepackage[symbol]{footmisc}
\usepackage{xr}
\newsavebox{\foobox}
\newcommand{\slantbox}[2][0]{\mbox{%
		\sbox{\foobox}{#2}%
		\hskip\wd\foobox
		\pdfsave
		\pdfsetmatrix{1 0 #1 1}%
		\llap{\usebox{\foobox}}%
		\pdfrestore
}}
\newcommand\unslant[2][-.25]{\slantbox[#1]{$#2$}}

\DeclareSIUnit{\micrometer}{\unslant{\mu}m}
\DeclareSIUnit{\microgram}{\unslant{\mu}\gram}
\DeclareSIUnit{\microM}{\unslant{\mu}M}
\DeclareSIUnit{\microliter}{\unslant{\mu}\liter}

\makeatletter
\let\cat@comma@active\@empty
\makeatother

\begin{document}

% The physical limits of constitutional sensing
% On the precision of constitutional sensing
% On the precision of sensing the constitution of random media
% On the precision of sensing constitutive relations
% The limits of sensing constitutive relations
% The limits of constitutional sensing

%Physical limits to sensing constitutive relations
%Physical limits to the perception of material properties
\title{Physical limits to sensing material properties}

\author{Farzan Beroz}
\email{Corresponding author. Email: farzan@umich.edu}
\affiliation{Department of Physics, University of Michigan, Ann Arbor, Michigan 48109, USA}
\author{Di Zhou}
\affiliation{Department of Physics, University of Michigan, Ann Arbor, Michigan 48109, USA}
\author{Xiaoming Mao}
\affiliation{Department of Physics, University of Michigan, Ann Arbor, Michigan 48109, USA}
\author{David K. Lubensky}
\affiliation{Department of Physics, University of Michigan, Ann Arbor, Michigan 48109, USA}

\begin{abstract} % abstract
Constitutive relations describe how materials respond to external stimuli such as forces. All materials respond heterogeneously at small scales, which limits what a localized sensor can discern about the global constitution of a material. In this paper, we quantify the limits of such constitutional sensing by determining the optimal measurement protocols for sensors embedded in disordered media. For an elastic medium, we find that the least fractional uncertainty with which a sensor can determine a material constant $\lambda_0$ is approximately

\begin{equation*}
 \frac{\delta \lambda_0}{\lambda_0 } \sim \left( \frac{\Delta_{\lambda} }{ \lambda_0^2} \right)^{1/2} \left( \frac{ d }{ a } \right)^{D/2} \left( \frac{ \xi }{ a } \right)^{D/2}   
\end{equation*}\\

\vspace*{-\baselineskip}

\noindent for $a \gg d \gg \xi$, $\lambda_0 \gg \Delta_{\lambda}^{1/2}$, and $D>1$, where $a$ is the size of the sensor, $d$ is its spatial resolution, $\xi$ is the correlation length of fluctuations in the material constant, $\Delta_{\lambda}$ is the local variability of the material constant, and $D$ is the dimension of the medium. Our results reveal how one can construct microscopic devices capable of sensing near these physical limits, e.g. for medical diagnostics. We show how our theoretical framework can be applied to an experimental system by estimating a bound on the precision of cellular mechanosensing in a biopolymer network.
\end{abstract}

\maketitle

\titleformat{\section}    
       {\normalfont\fontfamily{cmr}\fontsize{12}{17}\bfseries}{\thesection}{1em}{}
\titleformat{\subsection}[runin]
{\normalfont\fontfamily{cmr}\bfseries}{}{1em}{}

\renewcommand\thefigure{\arabic{figure}}    
%\renewcommand\thesection{Supplementary Note \arabic{section}:}

%\onecolumngrid

\setcounter{section}{0}
\setcounter{figure}{0}
\setcounter{equation}{0}

\setlength{\parskip}{0pt}

%\part*{\centerline{Summary}}

\section{Introduction}

A fundamental way of learning about a material is by observing how it responds to external stimuli. The functional dependence of a response on a stimulus is known as a constitutive relation. The most basic example of such a relation is Hooke's law $F=kX$ for the deformation response $X$ of a linear elastic solid to a force stimulus $F$, where $k$ is a material constant that is a characteristic property of the solid \cite{Hooke1678,Landau2004}. This linearity is a generic feature of material response for small enough stimuli, as it requires only that the constitutive relation be analytic and non-vanishing to first order. Linear constitutive relations have proven useful for characterizing a broad range of physical systems, including dielectric materials \cite{Hippel1995}, diffusion \cite{Fick1855}, friction \cite{Amontons1699}, geomaterials \cite{Darve2004}, Newtonian fluids \cite{Landau2013}, piezoelectric materials \cite{Curie1880}, thermoelectric materials \cite{Rowe2005}, and even abstract entities such as financial markets \cite{Iyetomi2011, Bouchaud2017}.

Material constants of linear constitutive relations are typically inferred by comparing the known value of an applied stimulus to the measured response produced by the stimulus. For the case of a homogeneous elastic solid, the material constant is simply given by $k = F/X$. In reality, however, almost all materials are spatially heterogeneous \cite{Ossi2002, DiDonna2005,Torquato2005, KurtBinder2011}. This heterogeneity serves as a source of measurement noise that becomes significant for systems that operate at the microscale, such as miniature electronic devices \cite{Fahlbuscha,Cullinan2012, JingW2014Real, Wallace2017}, medical microrobots \cite{Bhat2004,Nelson2006,Ornes2017,Simaan2018}, and biological sensors \cite{Discher2005,Arlett2011,Skedung2013,Yang2016,Doyle2016,Beroz2017,Petridou2017}.

Previous studies of sensing in random media have focused on remote sensing or communication via traveling waves \cite{Zuniga1980,Barbour1991,Kravtsov1993,Ishimaru1997,Moustakas2000,W.G.Rees2016}. The inference of material constants at small scales has been studied in microrheology \cite{Mason1995, Schnurr1997,Weihs2006} and for chemical sensing \cite{Berg1977,Bialek2005,Endres2008,Kaizu2014}. In these contexts, inference is typically performed by assuming homogeneity and exploiting thermal fluctuations. This type of passive sensing yields information about material constants that is bounded by fluctuation-dissipation theorems \cite{Kubo2, Bialek2005}. However, sensing in athermal systems calls for active forces. Although some methods are available to infer material constants using active probes \cite{Bausch1999, Helfer2000, Levine2000, Rigato2017}, the effect of spatial heterogeneity on this process has remained unclear \cite{DiDonna2005}. What are the theoretical limits to the precision of sensing in heterogeneous materials, and how can a physical device be designed to achieve these limits?

To quantify the limits of sensing constitutive relations, we investigate a simple model of a localized sensor interacting with a heterogeneous medium. Specifically, we consider a continuous medium with a material constant given by a uniform average value $\lambda_0$ plus a spatially-varying fluctuation $\delta \lambda(\boldsymbol{r})$ with short-ranged correlations. We treat the sensor as a spherical device that can probe $\lambda_0$ by applying an external stimulus field and measuring the resulting response field in equilibrium.

In what follows, we show that this inference process admits an optimal (minimum-variance unbiased) estimator of $\lambda_0$. The precision of this estimator depends on the form of the spatial response function of the medium. For a short-ranged response function, the precision is bounded because the sensor can only probe the material constant field in its immediate vicinity. A long-ranged response function enables the sensor to significantly improve its precision by accessing nonlocal information. Interestingly, however, this nonlocal information is subject to interference, and thus cannot be fully decoded using a single measurement. We demonstrate how a sensor can avoid this interference by performing a sequence of measurements with varying measurement protocols. This ``sensory multiplexing'' can increase the precision of a sensor by up to a factor proportional to a power of its spatial resolution. We conclude by using our theoretical framework to bound the precision of cellular mechanosensing in a biopolymer network, a sensory process known to regulate cellular behavior in decisive ways \cite{Doyle2016,Zaman2006,Guo2013,Thievessen2015}.

\section{Probing a Winkler foundation} \label{sec:probewinkler}

To gain insight into sensing constitutive relations in physical space, we explore a minimal theoretical model that consists of a spherical sensor embedded in a heterogeneous medium (see Fig. 1). In this section, we start by taking the medium to be the simplest heterogeneous material: a disordered Winkler foundation \cite{Winkler1868}. This medium corresponds to an array of decoupled springs in the continuum limit. The internal energy of the Winkler foundation is given by:

\begin{equation}
E = \frac{1}{2} \int \lambda(\boldsymbol{r}) u(\boldsymbol{r})^2  d\boldsymbol{r},
\end{equation}

\noindent where $\lambda(\boldsymbol{r})$ is a spatially-varying material constant and $u(\boldsymbol{r})$ is the response field at position $\boldsymbol{r}$. We assume $\lambda(\boldsymbol{r}) = \lambda_{0} + \delta \lambda(\boldsymbol{r})$, where $\lambda_{0}$ is a fixed, uniform field and $\delta \lambda(\boldsymbol{r}) \ll \lambda_0$ is a Gaussian random field with zero mean and spatial correlations given by:

\begin{equation}\label{eq:dldl}
\langle \delta \lambda(\boldsymbol{r}) \delta \lambda(\boldsymbol{r}') \rangle = \frac{\Delta_{\lambda}}{(2\pi)^{D/2}} e^{- (\boldsymbol{r} - \boldsymbol{r}')^2 / \xi^2},
\end{equation}

\noindent where $\Delta_{\lambda} \ll \lambda_0^2$ is the local variability of $\lambda(\boldsymbol{r})$, $D$ is the spatial dimension, and $\xi$ is the correlation length of the fluctuations in $\lambda(\boldsymbol{r})$. For simplicity, we assume $\xi$ is small enough that these correlations can be approximated by:

\begin{equation}\label{eq:dldla}
\langle \delta \lambda(\boldsymbol{r}) \delta \lambda(\boldsymbol{r}') \rangle = \Delta_{\lambda}\xi^D \delta (\boldsymbol{r} - \boldsymbol{r}').
\end{equation}

\noindent The quenched disorder $\delta \lambda(\boldsymbol{r})$ in the material constant limits the precision with which a physical sensor can infer $\lambda_{0}$. To determine these limits, we consider an idealized sensor that probes $\lambda_0$ by first applying a stimulus field $f(\boldsymbol{r})$. This field perturbs the energy of the system as follows:

\begin{equation}\label{eq:fspring}
\delta E = - \int f(\boldsymbol{r}) u(\boldsymbol{r}) d\boldsymbol{r}.
\end{equation}

\noindent After applying this stimulus, the sensor measures the response of the medium in equilibrium. In particular, we assume that the sensor records an integrated response $m$:

\begin{equation}\label{eq:mspring}
m = \int w(\boldsymbol{r}) u(\boldsymbol{r}) d\boldsymbol{r} ,
\end{equation}

\begin{figure}
	\centering
	\includegraphics[width=0.65\columnwidth]{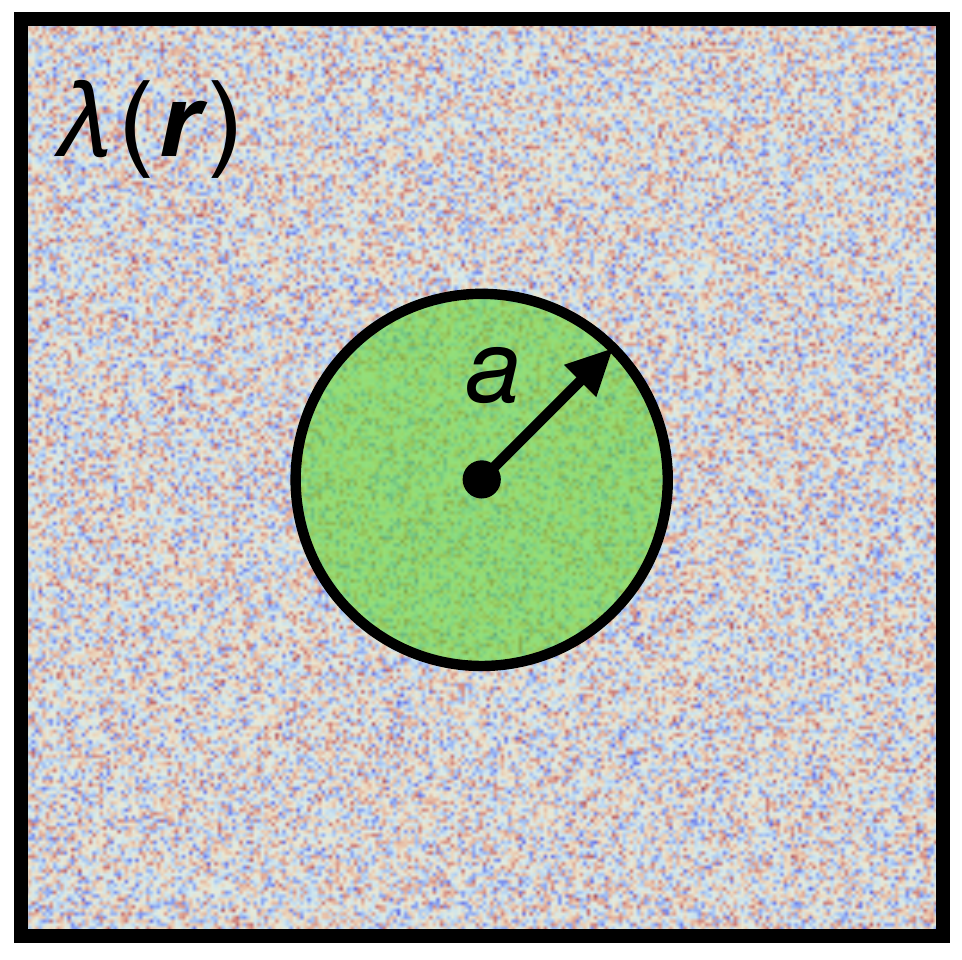}
	\caption{\label{fig:FG0}
		Sensing in a heterogeneous medium. Schematic illustration of sensing constitutive relations, showing an idealized, spherical sensor of radius $a$ (green) embedded inside a medium with a spatially-varying material constant field $\lambda(\boldsymbol{r})$ (background). The sensor can learn about $\lambda(\boldsymbol{r})$ by applying an arbitrary stimulus and recording an arbitrary weighted response within its volume.
	}
\end{figure}

\noindent where $w(\boldsymbol{r})$ is a weight field. Taken together, the probe fields $f(\boldsymbol{r})$ and $w(\boldsymbol{r})$ define the measurement protocol of the sensor. For any physical sensor, these fields must be localized in space. We impose this locality by constraining the probe fields to obey $f(\boldsymbol{r}) = 0$ and $w(\boldsymbol{r}) = 0$ for $r>a$, where $r$ is the radial coordinate and $a$ is the radius of the sensor.

Finally, upon recording the integrated response $m$, the sensor produces an estimate for $\lambda_0$. In what follows, we will determine the optimal estimator $\hat{\lambda}_0$ perturbatively to leading order in $\delta \lambda (\boldsymbol{r})$. In this approximation, the integrated response is:

\begin{equation} \label{eq:mspringint}
m = \int \left( \frac{1}{\lambda_0} -  \frac{\delta \lambda (\boldsymbol{r}) }{\lambda_0^2} \right) \psi(\boldsymbol{r}) d\boldsymbol{r},
\end{equation}

\noindent where we have defined the probe intensity $\psi(\boldsymbol{r}) \equiv f(\boldsymbol{r})w(\boldsymbol{r})$. For a fixed choice of $\psi(\boldsymbol{r})$, along with prior knowledge of the model parameters other than $\lambda_0$, the optimal estimator of $\lambda_0$ based on the outcome of $m$ is (see Supplemental Material, Sec. \ref{sec:optwink}):

\begin{equation}\label{eq:haha}
\hat{\lambda}_0 = \frac{s}{m},
\end{equation}

\noindent where $s$ is a normalizing constant chosen such that $\hat{\lambda}_0$ yields an unbiased estimate of $\lambda_0$:

\begin{equation}\label{eq:sspring}
s = \int \psi(\boldsymbol{r}) d\boldsymbol{r}.
\end{equation}

\noindent Eq. (\ref{eq:haha}) is a mesoscopic generalization of Hooke's law $k=F/X$. By computing the estimate $\hat{\lambda}_0$, the sensor obtains a weighted spatial average of $\lambda(\boldsymbol{r})$:

\begin{equation} \label{eq:weightedavg}
\hat{\lambda}_0 = \frac{\int \psi(\boldsymbol{r}) \lambda(\boldsymbol{r}) d\boldsymbol{r} }{ \int \psi(\boldsymbol{r}) d\boldsymbol{r}},
\end{equation}

\noindent to leading order in $\delta\lambda(\boldsymbol{r})$. This estimator is optimal in the sense that it has a lower variance $\delta \lambda_0^2 \equiv \langle (\hat{\lambda}_0 - \lambda_0)^2 \rangle$ than any other unbiased estimator for a fixed choice of measurement protocol. Therefore, the optimal measurement protocol can be determined by minimizing $\delta\lambda_0^2$ with respect to the probe intensity $\psi(\boldsymbol{r})$. Inserting Eq. (\ref{eq:dldla}) into the definition of the variance yields:

\begin{equation} \label{eq:varspring}
\delta\lambda_0^2 =  \Delta_{\lambda} \xi^{D} \frac{ \int \psi(\boldsymbol{r})^2 d\boldsymbol{r}}{  \left( \int \psi(\boldsymbol{r}) d\boldsymbol{r} \right)^2}.
\end{equation}

\noindent This variance is invariant with respect to an overall rescaling of $\psi(\boldsymbol{r})$. To eliminate this redundancy, we constrain $\int \psi(\boldsymbol{r}) d\boldsymbol{r}$ to be a fixed constant. Furthermore, we must enforce $\psi(\boldsymbol{r}) = 0$ in the exterior of the sensor ($r>a$) to satisfy the constraints imposed by the finite size of the sensor. Thus, the minimum of $\delta \lambda_0^2$ is determined by the configuration of $\psi(\boldsymbol{r})$ that extremizes the following action $S$:

\begin{equation} \label{eq:aa}
S = \int_{\mathcal{R}_{\mathrm{int}}} \left(  \frac{1}{2}\psi(\boldsymbol{r})^2 - \gamma \psi(\boldsymbol{r}) \right) d\boldsymbol{r},
\end{equation}

\noindent where the integral is taken over the interior $\mathcal{R}_{\mathrm{int}}$ of the sensor ($r<a$) and $\gamma$ is a Lagrange multiplier that fixes $\int \psi(\boldsymbol{r})d\boldsymbol{r}$. This action is extremized by any measurement protocol with a probe intensity $\psi(\boldsymbol{r})$ that is uniform over $\mathcal{R}_{\mathrm{int}}$. The optimal measurement protocol is therefore:

\begin{equation} \label{eq:psiint}
\psi(\boldsymbol{r})  =\begin{cases}
\gamma, & \text{$r < a$}.\\
0, & \text{$r>a$}.
\end{cases}
\end{equation}

\noindent Inserting Eq. (\ref{eq:psiint}) into Eq. (\ref{eq:varspring}) yields:

\begin{equation} \label{eq:winkvar}
\delta\lambda_0^2 = \Delta_{\lambda} \xi^{D} V^{-1},
\end{equation}

\noindent where $V$ is the volume of the sensor. Thus, the fractional uncertainty of the estimator $\hat{\lambda}_0$, defined as the standard deviation $\delta \lambda_0$ divided by the mean $\lambda_0$, scales as:

\begin{equation} \label{eq:fuwinkler}
 \frac{\delta \lambda_0}{\lambda_0 } \sim \left( \frac{\Delta_{\lambda} }{ \lambda_0^2} \right)^{1/2}   \left( \frac{ \xi }{ a } \right)^{D/2},
\end{equation}

\noindent which can be interpreted as the familiar $1/\sqrt{N}$ scaling of measurement uncertainty for $N$ independent samples. In this analogy, the sample size $N\sim (a/\xi)^{D}$ corresponds to the number of effectively independent subvolumes probed by the sensor.

\section{Probing an elastic sheet with dipoles} \label{sec:probedipole}

%Increasing the range of the response function enables more accurate sensing
%The accuracy of inference depends on the range of the response functio

For the Winkler foundation, our model sensor could not induce a response beyond its volume. In contrast, many other types of elastic media are coupled in space and thereby respond to stimuli nonlocally. To understand how such nonlocality affects a sensor's ability to infer material properties, we now turn to conventional, linear elasticity. For simplicity, we will first focus on an isotropic, two-dimensional elastic sheet characterized by a single material constant, and in Sec. \ref{sec:mechanosensing} we will generalize our theoretical framework to a three-dimensional elastic medium characterized by a material constant tensor.

For the elastic sheet, we consider the deformation response $u(\boldsymbol{r})$ to force stimuli $f(\boldsymbol{r})$ oriented perpendicular to the plane of the sheet. Thus, the sheet's internal energy depends on the gradient $\nabla u(\boldsymbol{r})$ of the response field as follows:

\begin{equation} \label{eq:esheet}
E = \frac{1}{2} \int \lambda(\boldsymbol{r}) \nabla u(\boldsymbol{r}) \cdot \nabla  u(\boldsymbol{r}) d\boldsymbol{r}.
\end{equation}

\noindent Here, as in the previous section, we take $\lambda(\boldsymbol{r})$ to be a Gaussian random field with mean $\lambda_0$, variance $\Delta_{\lambda} \ll \lambda_0^2$, and spatial correlations over a scale $\xi$. As before, we take the sensor to interact with the medium within a radius $a$ by first applying a stimulus field $f(\boldsymbol{r})$ as in Eq. (\ref{eq:fspring}), and then measuring an integrated response $m$ as in Eq. (\ref{eq:mspring}).

To leading order in $\delta \lambda(\boldsymbol{r})$, the sensor can again compute $\hat{\lambda}_0 = s/m$ to obtain a spatial average of $\lambda(\boldsymbol{r})$ weighted by a probe intensity $\psi(\boldsymbol{r})$, as in Eq. (\ref{eq:weightedavg}) (see Supplemental Material, Sec. \ref{sec:pies}). However, for the elastic sheet, $\psi(\boldsymbol{r})$ is now:

\begin{equation} \label{eq:psisheet}
\psi(\boldsymbol{r}) = \nabla V_f(\boldsymbol{r}) \cdot \nabla V_w(\boldsymbol{r}),
\end{equation}

\noindent where $V_f( \boldsymbol{r} )$ and $V_w( \boldsymbol{r} )$ are scalar potentials associated with the probe fields $f( \boldsymbol{r} )$ and $w( \boldsymbol{r} )$:

\begin{equation}
V_f( \boldsymbol{r} ) = \int G(\boldsymbol{r} - \boldsymbol{r'}) f(\boldsymbol{r'}) d\boldsymbol{r'},
\end{equation}

\begin{equation}
V_w( \boldsymbol{r} ) = \int G(\boldsymbol{r} - \boldsymbol{r'}) w(\boldsymbol{r'}) d\boldsymbol{r'},
\end{equation}

\noindent which are assumed to be continuous and to vanish at infinity. Here, $G(\boldsymbol{r} - \boldsymbol{r}') = \ln{|\boldsymbol{r}-\boldsymbol{r}'|}/(2\pi)$ is the response function for a homogeneous sheet, i.e. the solution of $\nabla^2 G(\boldsymbol{r}-\boldsymbol{r'}) = \delta (\boldsymbol{r}-\boldsymbol{r'})$. This response function is long-ranged, and thereby allows the sensor to probe distant regions beyond its boundary.

Intuitively, probing a greater extent of the medium should yield a more accurate estimate of $\lambda_0$. To that end, the greatest possible extent of a probe is achieved by probe potentials with a $\sim 1/r$ radial dependence in the far-field limit. For the elastic sheet, this decay profile is not produced by monopoles (which yield pathological, non-decaying potentials), but rather by dipoles. The simplest possible measurement protocol with dipole probe fields is described by:

\begin{equation} \label{eq:dphase1}
f(\boldsymbol{r}) \sim \delta(r - a) \cos(\theta),
\end{equation}

\begin{equation} \label{eq:dphase2}
w(\boldsymbol{r}) \sim \delta(r - a) \cos(\theta).
\end{equation}

\noindent These probe fields cast a probe intensity $\psi(\boldsymbol{r})$ that is uniform in the interior of the sensor and isotropically decaying in the exterior:

\begin{equation} \label{eq:psiint2}
\psi(\boldsymbol{r})  =\begin{cases}
\gamma, & \text{$r<a$}.\\
\gamma \left( \frac{a}{r} \right)^4, & \text{$r>a$}.
\end{cases}
\end{equation}

\noindent Inserting Eq. (\ref{eq:psiint2}) into Eq. (\ref{eq:varspring}) yields the following variance:

\begin{equation}
\delta \lambda_0^2 = \frac{1}{3}\Delta_{\lambda} \xi^{D} V^{-1},
\end{equation}

\noindent for $D=2$. As expected from dimensional analysis, this expression has the same dependence on the model parameters as for the Winkler foundation (cf. Eq. (\ref{eq:winkvar})). Importantly, however, its prefactor is smaller. Thus, our example illustrates how a sensor can harness a long-ranged response function to perform at a higher precision by effectively averaging $\lambda(\boldsymbol{r})$ over a larger region of space.

\section{Probe field interference limits the channel capacity of sensing} \label{sec:probeinter}

\begin{figure*}
	\centering
	\includegraphics[width=1.5\columnwidth]{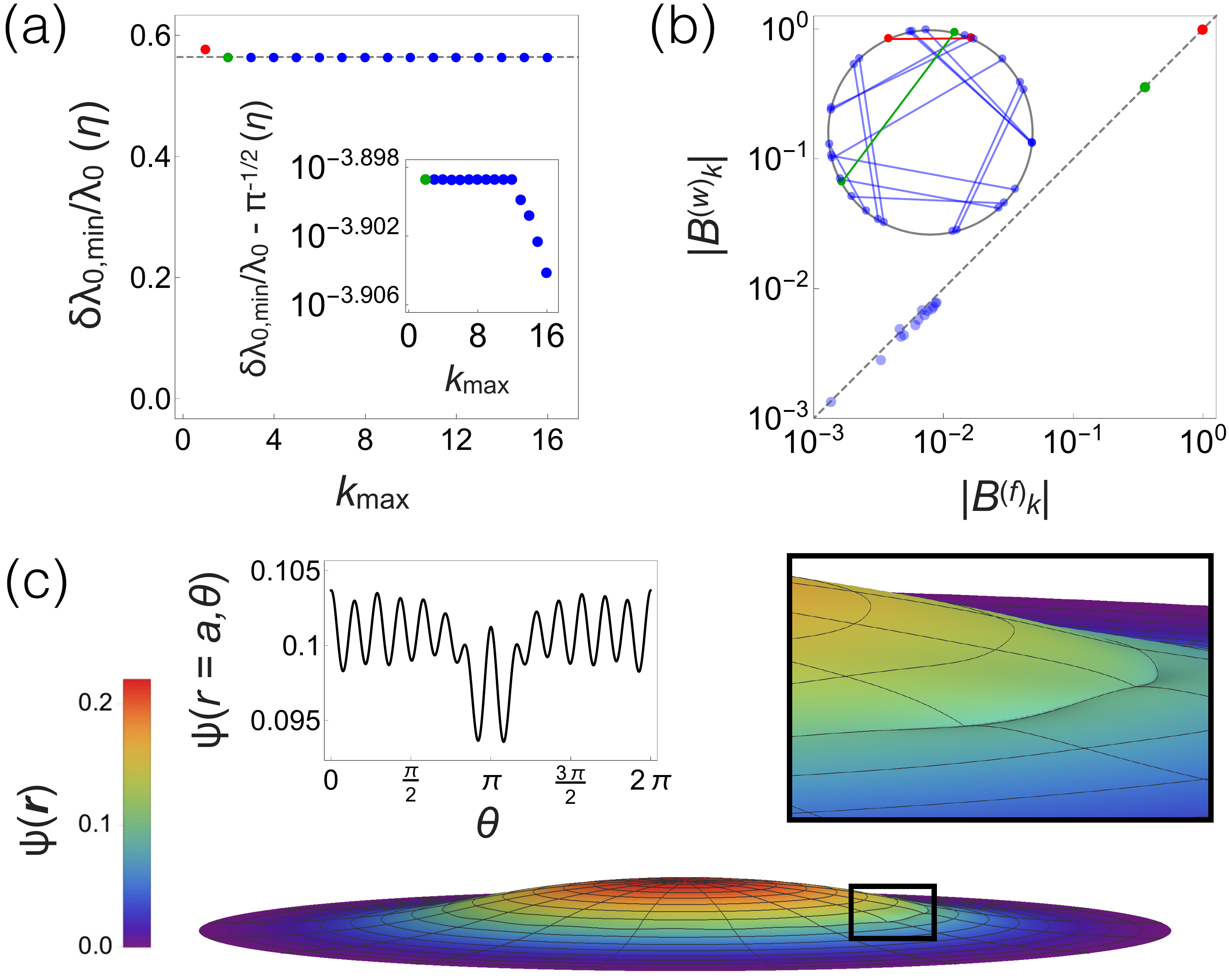}
	\caption{\label{fig:FG0}
		Probe field interference limits the amount of information that a sensor can glean from a single measurement. (a) Fractional uncertainty $\delta \lambda_{0,\mathrm{min}}/\lambda_0$ in units of $\eta = \xi^D \Delta_{\lambda} V^{-1}$ for numerically-optimal boundary probes versus maximum absolute mode number $k_{\mathrm{max}}$ included in the multipole expansions of the probe fields (see Sec. \ref{sec:probeinter}). Red point corresponds to the dipole-dipole measurement protocol ($k_{\mathrm{max}}=1$) considered in Sec. \ref{sec:probedipole}, green point to $k_{\mathrm{max}}=2$, and blue points to $k_{\mathrm{max}}>2$. Dashed gray line indicates the optimum $\delta \lambda_{0,\mathrm{low}}/\lambda_0 \approx \eta / \sqrt{\pi} $ attained in the limit $k_{\mathrm{max}}\rightarrow \infty$ for the convex relaxation of $\delta\lambda_0 / \lambda_0$ described in Sec. \ref{sec:probeinter}. Inset: $\delta \lambda_{0,\mathrm{min}}/\lambda_0 - 1 / \sqrt{\pi}$ in units of $\eta$ versus $k_{\mathrm{max}}$ on a logarithmic scale. (b) Absolute values of the weight field coefficients $|B^{(w)}_k|$ versus the absolute values of the stimulus field coefficients $|B^{(f)}_k|$ for an example measurement protocol obtained via numerical optimization for $k_{\mathrm{max}} = 16$, showing dipole modes (red), quadrupole modes (green), and higher order modes (blue). Dashed gray line shows $|B^{(w)}_k| = |B^{(f)}_k|$. Inset: phases of the probe field coefficients for the same example measurement protocol as in the main panel. Lines connect the coefficients of the stimulus and weight fields that correspond to the same value of $k$. Colors same as in main panel. (c) Same example probe intensity $\psi(\boldsymbol{r})$ as in panel (b) versus spatial coordinate $\boldsymbol{r}$. Left inset: $\psi(\boldsymbol{r})$ at the boundary of the sensor ($r=a$) versus angular coordinate $\theta$. Right inset: larger view of the region indicated by the black rectangle in the main panel, showing small wrinkles in $\psi(\boldsymbol{r})$.
	}
\end{figure*}

The dipole-dipole measurement protocol we considered above for the elastic sheet achieves a fractional uncertainty of $\delta \lambda_0 / \lambda_0 = \eta / \sqrt{3}$, where $\eta \equiv \lambda_{0}^{-1\vphantom{/2}}\Delta_{\lambda}^{1/2}\xi^{D/2}_{\vphantom{0}}V^{-1/2}_{\vphantom{0}}$ is the smallest possible fractional uncertainty for the Winkler foundation in two dimensions. Given that a probe of the elastic sheet can access nonlocal information, what limits its precision? To answer this question, we start by considering a sensor that can apply an arbitrary pair of probe fields on its boundary. For such boundary probes, the most general probe fields are of the form:

\begin{equation} \label{eq:dphasef1}
f(\boldsymbol{r}) \sim \delta(r - a) \sum_{k}  B^{(f)}_k e^{i k\theta} ,
\end{equation}

\begin{equation} \label{eq:dphasef2}
w(\boldsymbol{r}) \sim \delta(r - a) \sum_{k} B^{(w)}_k e^{- i k\theta},
\end{equation}

\noindent where $B^{(f)}_k$ and $B^{(w)}_k$ are complex coefficients that satisfy $B^{(f)}_{-k} = B^{(f)*}_k$ and $B^{(w)}_{-k} = B^{(w)*}_k$ to ensure that the probe fields are real. This measurement protocol gives rise to the following probe intensity:

\begin{equation} \label{eq:beats}
\psi_{\pm}(\boldsymbol{r}) = \sum_{k,l}  B_{kl}  (kl +  |k| | l | )   \left( \frac{r}{a} \right)^{ \pm |k| \pm |l| - 2} e^{i (k - l) \theta},
\end{equation}

\noindent where $\psi_{+}(\boldsymbol{r})$ and $\psi_{-}(\boldsymbol{r})$ correspond to the interior ($r<a$) and the exterior ($r>a$) of the sensor, respectively, and $B_{kl} \sim B^{(f)}_{k} B^{(w)}_{l}$. From this expression, we see that the isotropy of $\psi(\boldsymbol{r})$ in Eq. (\ref{eq:psiint2}) is a general feature of measurement protocols that consist of pure multipoles of equal mode number. In contrast, discordant modes generically ``interfere'' to yield probe intensities that beat as a function of the angular coordinate.

Inserting this probe intensity into Eq. (\ref{eq:varspring}) and performing the integrals yields the following variance:

\begin{equation} \label{eq:tensorone}
\delta \lambda_0^2 = \Delta_{\lambda} \xi^{D} \sum_{k,l,m,n}  B_{kl} B_{mn} \mathbb{T}_{klmn},
\end{equation}

\noindent where $\mathbb{T}_{klmn}$ is a highly structured, fourth-order tensor:

\begin{equation}
\mathbb{T}_{klmn} = 4\pi \frac{ \delta_{k-l + m - n, 0} x_{klmn} y_{klmn} }{( x_{klmn} + 2 )(x_{klmn} - 2)}.
\end{equation}

\noindent Here, $\delta_{i,j}$ is the Kronecker delta function, $x_{klmn} = \left|k\right|+\left| l \right| + \left|m \right| + \left|n \right|$, and $y_{klmn} = (\left|k \right| + k + \left| l \right| + l)(\left|m \right| + m + \left|n \right| + n)$. In Eq. (\ref{eq:tensorone}), we have fixed the normalizing constant to be $\int \psi(\boldsymbol{r}) d\boldsymbol{r} = 1$, which implies that the coefficients $B_{kl}$ must obey:

\begin{equation} \label{eq:tensortwo}
\sum_k 2\pi |k| B_{kk} = 1.
\end{equation}

\noindent To gain insight into the optimal measurement protocols for this sensory geometry, we numerically minimized the fractional uncertainty $\delta \lambda_0 / \lambda_0$. To that end, we imposed a cutoff on the system by truncating the sums in Eqs. (\ref{eq:tensorone}) and (\ref{eq:tensortwo}) at a maximum absolute mode number $k_{\mathrm{max}}$. Physically, this parameter corresponds to the spatial resolution of the sensor, which we define as:

\begin{equation} \label{eq:resolution}
d = \left(\frac{2\pi }{k_{\mathrm{max}}} \right)a.
\end{equation}

\noindent For a given value of $k_{\mathrm{max}}$, we determined the minimum fractional uncertainty $\delta \lambda_{0,\mathrm{min}} / \lambda_0$ by using the Nelder-Mead algorithm to optimize over the values of the coefficients $B^{(f)}_k$ and $B^{(w)}_k$ (see Supplemental Material, Sec. \ref{sec:snum}). For all choices of $k_{\mathrm{max}}$ we studied, this algorithm consistently converged to basins of minima dominated by the dipole modes, as we intuited in Sec. \ref{sec:probedipole}. Interestingly, however, as we increased $k_{\mathrm{max}}$, we found that at certain special values, the optimal probe fields shifted and picked up additional higher order modes, resulting in a smaller fractional uncertainty $\delta \lambda_{0,\mathrm{min}}/\lambda_0$ (see Fig. 2(a)). The higher order modes contribute with smaller amplitudes and nontrivial relative phase shifts (see Fig. 2(b)). These complex configurations arise because different terms in Eq. (\ref{eq:tensorone}) can provide conflicting contributions to the variance depending on the relative phases of the modes. This geometrical frustration greatly suppresses the inclusion of modes beyond the dipole-dipole and quadrupole-quadrupole pairs, which for $2 \le k_{\mathrm{max}} \le 12$ appear together with phase relations that result in an isotropic $\psi(\boldsymbol{r})$. Including three or more mode pairs must break isotropy in a manner that is analogous to the impossibility of simultaneously minimizing the interaction energies among three or more antiferromagnetically interacting spins (see Supplemental Material Sec. \ref{sec:breakisotropy}). Nevertheless, for $k_{\mathrm{max}} > 12$, the optimal measurement protocols contain additional higher order modes that cause small wrinkles in $\psi(\boldsymbol{r})$ (see Fig. 2(c)). Although these wrinkles break the isotropy of $\psi(\boldsymbol{r})$, they also smoothen out its profile in the radial direction, which results in a greater overall uniformity throughout space and thus a higher precision.

To better understand the asymptotic behavior of the optimal measurement protocols for large $k_{\mathrm{max}}$, we imagine relaxing the constraints on the coefficients $B^{(f)}_k$ and $B^{(w)}_k$ by allowing $B_{kl}$ in Eqs. (\ref{eq:tensorone}) and (\ref{eq:tensortwo}) to be an arbitrary matrix satisfying $B_{-k,-l} = B_{kl}^*$. This relaxation expands the space of possible $\psi(\boldsymbol{r})$ to include all real configurations that can be generated by Eq. (\ref{eq:beats}), some of which cannot be cast by a physical probe. Importantly, this relaxation is a convex function of the coefficient matrix $B_{kl}$, and thus has a unique minimum $\delta \lambda_{0,\mathrm{low}}/\lambda_0$ that provides a theoretical lower bound on $\delta\lambda_0 / \lambda_0$. Specifically, in the limit $k_{\mathrm{max}} \rightarrow \infty$, we find that $\delta \lambda_{0,\mathrm{low}} / \lambda_0 \approx \eta / \sqrt{\pi}$, which provides a close lower bound on the values obtained via numerical minimization (see Fig. 2(a) and Supplemental Material, Sec. \ref{sec:convexrelax}).

A simple argument based on symmetry reveals that this lower bound must be an inequality for $k_{\mathrm{max}} > 2$. This argument follows from observing that for all values of $k_{\mathrm{max}}$, the unique optimal configuration of $\psi(\boldsymbol{r})$ for the relaxation is isotropic, in contrast to the optimal configurations we found by numerically minimizing Eq. (\ref{eq:tensorone}) for $k_{\mathrm{max}} > 12$ (see Supplemental Material, Sec. \ref{sec:convexrelax}). This broken isotropy must persist for all higher values of $k_{\mathrm{max}}$, and therefore a boundary probe can never cast a configuration of $\psi(\boldsymbol{r})$ that performs as well as the optimal $\psi(\boldsymbol{r})$ for the convex relaxation of $\delta\lambda_0 / \lambda_0$. This example illustrates how interferences between the probe fields fundamentally limit the amount of information that can be gleaned from a single measurement, i.e. the channel capacity of sensing. In the following section, we will show how a sensor can overcome this limit by performing multiple probes, and then we will generalize our results to a sensor that can apply arbitrary probe fields within its volume.

\section{Sensory multiplexing can significantly improve the precision of sensing} \label{sec:greatjob}

In the previous section, we found that the probe fields interfere to limit the precision of sensing. These interferences occur because all of the modes contained in the probe fields interrogate the medium simultaneously. In principle, however, each mode couples to a different spatial extent of the medium and therefore should carry independent information about $\lambda_{0}$. Such information could potentially be accessed by performing separate measurements with distinct spectra.

To test this notion, we determine the optimal estimator of $\lambda_0$ for a sensor that can perform multiple probes at a fixed location with varying measurement protocols. For concreteness, we label each probe by an integer $k$, and we constrain their probe fields $f_k(\boldsymbol{r})$ and $w_k(\boldsymbol{r})$ to be zero for $r>a$. In this case, the minimum-variance unbiased estimator of $\lambda_0$ is again given by a weighted spatial average of $\lambda(\boldsymbol{r})$:

\begin{equation}  \label{eq:psicoefs1}
\hat{\lambda}_0 = \frac{ \int \Psi(\boldsymbol{r}) \lambda(\boldsymbol{r}) d\boldsymbol{r}}{  \int \Psi(\boldsymbol{r}) d\boldsymbol{r}   }.
\end{equation}

\noindent Here, $\Psi(\boldsymbol{r})$ is an effective probe intensity created by the optimally weighted sum of the probe intensities $\psi_k(\boldsymbol{r})$ for the individual probes:

\begin{equation} \label{eq:psicoefs2}
\Psi(\boldsymbol{r}) = \sum_{k} p_k \frac{\psi_k(\boldsymbol{r})}{\int \psi_k(\boldsymbol{r}) d\boldsymbol{r}},
\end{equation}

\noindent where $p_k = \sum_{l} C^{-1}_{kl}$ with $C_{kl} \equiv \langle (\hat{\lambda}_{0,k} - \lambda_0) (\hat{\lambda}_{0,l} - \lambda_0) \rangle$ defined as the covariance matrix of the estimators $\hat{\lambda}_{0,k}$ for the individual probes (see Supplemental Material, Sec. \ref{sec:optmult}). The variance of the estimator $\hat{\lambda}_0$ is:

\begin{equation} \label{eq:cvar}
\delta \lambda_0^2 = \left( \sum_{k,l} {C}_{kl}^{-1} \right)^{-1}.
\end{equation}

\noindent According to Eqs. (\ref{eq:psicoefs1}) and (\ref{eq:psicoefs2}), $\hat{\lambda}_0$ is bilinear in $f_k(\boldsymbol{r})$ and $w_k(\boldsymbol{r})$. This bilinearity implies that is not possible to extract additional information by varying only one of the two probe fields of an optimal measurement protocol (see Supplemental Material, Sec. \ref{sec:eqprot}). Therefore, we consider a sensor that can vary the stimulus field and the weight field together as follows:

\begin{equation} \label{eq:multf}
f_k(\boldsymbol{r}) \sim \delta(r - a) \cos(k \theta),
\end{equation}

\begin{equation} \label{eq:multw}
w_k(\boldsymbol{r})  \sim \delta(r - a) \cos(k \theta).
\end{equation}

\noindent By varying the angular distributions of the probe fields in this manner, the sensor modulates its effective range in the exterior at the cost of simultaneously modulating the probe intensity $\psi_k(\boldsymbol{r})$ in the interior:

\begin{equation} \label{eq:psiint4}
\psi_k(\boldsymbol{r})  \sim \begin{cases}
\left( \frac{r}{a} \right)^{2k-2}, & \text{$r<a$}.\\
\left( \frac{r}{a} \right)^{-2k-2}  , & \text{$r>a$}.
\end{cases}
\end{equation}

\noindent To be concrete, we assume that the sensor executes a series of such probes from an initial mode number $k=1$ up to a maximum mode number $k=k_{\mathrm{max}}$, which corresponds to the spatial resolution $d$ of the sensor defined by Eq. (\ref{eq:resolution}) in Sec. \ref{sec:probeinter}.

\begin{figure*}
	\centering
	\includegraphics[width=1.5\columnwidth]{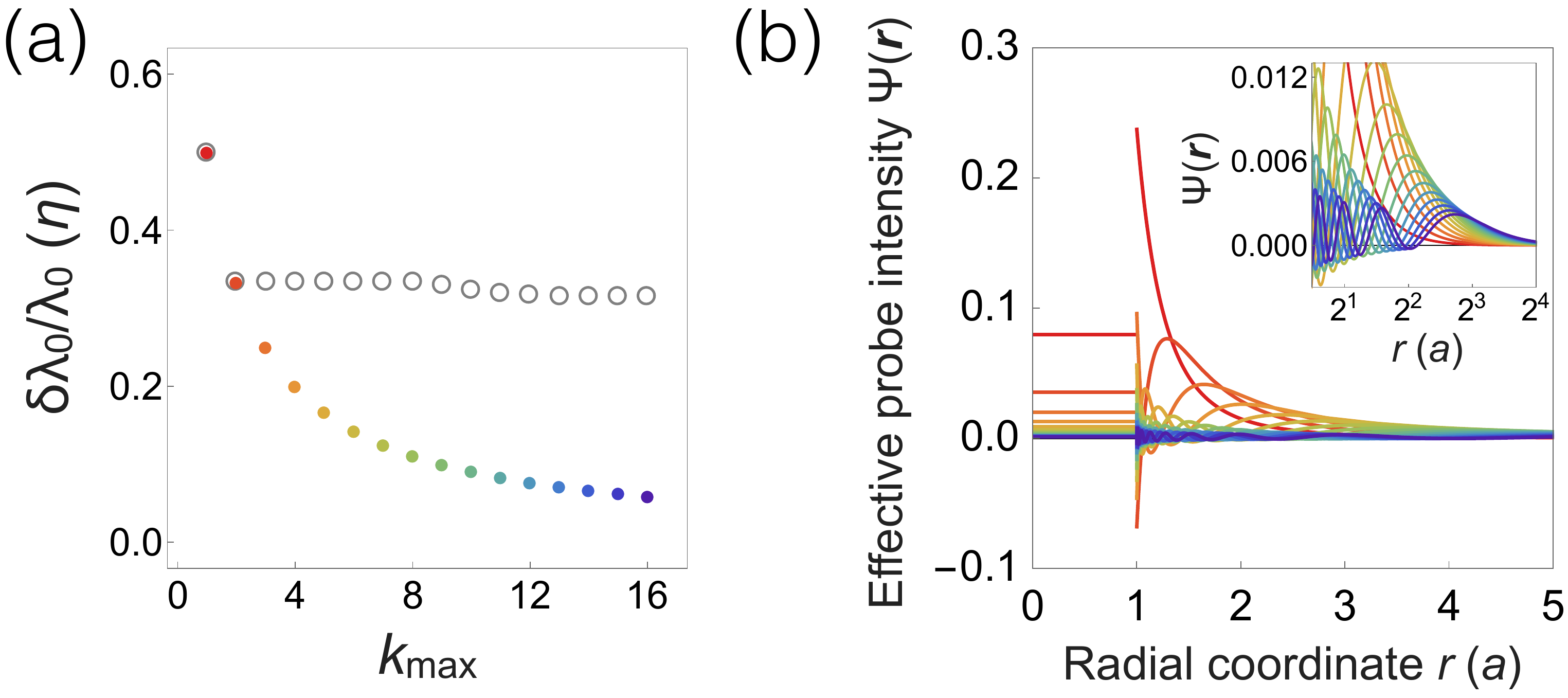}
	\caption{\label{fig:FG0}
		Sensory multiplexing can greatly improve the precision of sensing. (a) Colored points show smallest attainable fractional uncertainty $\delta \lambda_{0}/\lambda_0$ in units of $\eta = \xi^D \Delta_{\lambda} V^{-1}$ for a sensor that can perform sensory multiplexing up to a maximum absolute mode number $k_{\mathrm{max}}$ (see Sec. \ref{sec:greatjob}). Gray circles show a lower bound $\delta \lambda_{0,\mathrm{low}}/\lambda_0$ on the fractional uncertainty for each value of $k_{\mathrm{max}}$ for a single volume probe, obtained via numerical minimization (see Supplemental Material, Sec. \ref{sec:bulknum}). (b) Optimal effective probe intensities $\Psi(\boldsymbol{r})$ for sensory multiplexing versus radial coordinate $r$ in units of the sensor radius $a$ for the same values of $k_{\mathrm{max}}$ as in panel (a) (correspondence indicated by matching colors). Inset shows a larger view of the radial profiles of $\Psi(\boldsymbol{r})$ in the exterior plotted on a log-linear scale.
	}
\end{figure*}

Interestingly, this collection of boundary probes does not achieve a significant improvement over the dipole-dipole protocol we considered in Sec. \ref{sec:probedipole}. Instead, as $k_{\mathrm{max}}$ is increased, the fractional uncertainty approaches $ \delta \lambda_0 / \lambda_0 \approx \eta / \sqrt{\pi}$, as we found for the convex relaxation in Sec. \ref{sec:probeinter}. This agreement is not a mere coincidence: for boundary probes, the possible configurations of $\Psi(\boldsymbol{r})$ are mathematically equivalent to the possible configurations of $\psi(\boldsymbol{r})$ for the convex relaxation of a single probe (see Supplemental Material, Sec. \ref{sec:smsheet}). However, unlike the convex relaxation, the case of sensory multiplexing reveals an additional physical effect that can limit the precision of a sensor. That is, for multiple probes, the overlapping configurations of $\psi_k(\boldsymbol{r})$ in the interior correlate the probes and thereby suppress the amount of information that can be extracted from the exterior. These correlations are reflected in the structure of the covariance matrix, which is given by:

\begin{equation} \label{eq:covmaintext}
C_{kl} =  \frac{1}{4} \Delta_{\lambda} \xi^{2} V^{-1}   \left( \frac{k l }{k + l -1 } + \frac{k l}{k + l + 1} \right).
\end{equation}

\noindent In this expression, the first and second fractions are contributed by overlaps in the interior and exterior, respectively. To compensate for the superfluous contributions from the interior, the sensor must employ probe fields that are nonzero within its volume. One way to perform this compensation is by pairing each probe $k$ with a companion probe described by:

\begin{equation} \label{eq:companion1}
\tilde{V}_{f,k}(\boldsymbol{r}) \sim r-a,
\end{equation}

\begin{equation} \label{eq:companion2}
\tilde{V}_{w,k}(\boldsymbol{r}) \sim \left(\frac{r}{a}\right)^{2k-1} - 1,
\end{equation}

\noindent for $r<a$ and $0$ otherwise. These companion probe potentials result in probe intensities $\tilde{\psi}_k(\boldsymbol{r})$ that are confined to the interior of the sensor:

\begin{equation} \label{eq:psiint5}
\tilde{\psi}_k(\boldsymbol{r})  \sim \begin{cases}
\left( \frac{r}{a} \right)^{2k-2}, & \text{$r<a$}.\\
0, & \text{$r>a$}.
\end{cases}
\end{equation}

\noindent Pairing these companion probes with the original probes using Eq. (\ref{eq:psicoefs2}) with appropriate values of $p_k$ yields effective probe intensities $\Psi_k(\boldsymbol{r})$ that are zero in the interior (see Supplemental Material Sec. \ref{sec:smsheet}):

\begin{equation} \label{eq:pairedprobepsimt}
\Psi_k(\boldsymbol{r})  \sim \begin{cases}
0, & \text{$r<a$}.\\
 \left(\frac{r}{a}\right)^{-2k-2}, & \text{$r>a$}.
\end{cases}
\end{equation}

\noindent Finally, the sensor may include an additional unpaired probe $\psi_0(\boldsymbol{r})$ with a probe intensity given by Eq. (\ref{eq:psiint}) to uniformly sample the material constant field in its interior. With these adjustments, the resulting all-inclusive effective probe intensity $\Psi(\boldsymbol{r})$ exhaustively decodes the information available to the sensor (see Supplemental Material Sec. \ref{sec:smproof}). In this case, the covariances among the paired probes $\Psi_k(\boldsymbol{r})$ and the unpaired probe $\psi_0(\boldsymbol{r})$ are given by:

\begin{equation} \label{eq:cmnsol}
C_{kl} =  \Delta_{\lambda} \xi^{2} V^{-1}   \left( \delta_{k,0}\delta_{l,0} + \frac{k l}{k + l + 1} \right),
\end{equation}

\noindent for $k,l \ge 0$, where $k,l=0$ correspond to the unpaired probe. Inserting the inverse of this matrix into Eq. (\ref{eq:cvar}) yields the following variance:

\begin{equation} \label{eq:thefundamentalbound}
\delta \lambda_0^2 = \Delta_{\lambda} \xi^2 V^{-1}  \left(\frac{1}{k_{\mathrm{max}} + 1}  \right)^2.
\end{equation}

\noindent This variance decreases with $k_{\mathrm{max}}$ because each additional probe increases the uniformity of $\Psi(\boldsymbol{r})$ over space (see Fig. 3). In the limit of very fine resolution $k_{\mathrm{max}} \gg 1$ ($d\ll a$), the fractional uncertainty of the sensor's estimate of $\lambda_0$ scales as:

\begin{equation} \label{eq:funcert}
\frac{\delta \lambda_0}{\lambda_0 } \sim \left( \frac{\Delta_{\lambda} }{ \lambda_0^2} \right)^{1/2} \left( \frac{ d }{ a } \right)^{D/2} \left( \frac{ \xi }{ a } \right)^{D/2} ,
\end{equation}

\noindent for $D=2$. Thus, simultaneously varying both probe fields throughout the volume of the sensor can allow a significant amount of additional information to be transmitted across the sensory channel. We refer to this strategy as ``sensory multiplexing.''

Sensory multiplexing can be generalized to a three-dimensional elastic medium by taking the probe potentials to be pairs of spherical harmonics. In this case, the fractional uncertainty obeys the asymptotic scaling in Eq. (\ref{eq:funcert}) with $D=3$ (see Supplemental Material, Sec. \ref{sec:smsolid}). Moreover, the scaling is robust to the omission of a finite number of modes (see Supplemental Material, Sec. \ref{sec:smomit}). Taken together, our results reveal that for $D>1$ and $d \ll a$, sensory multiplexing can increase the effective volume sampled by a sensor by a factor proportional to the number $(a/d)^D$ of distinct subvolumes that it can resolve simultaneously.

Interestingly, this level of precision can never be attained by a single probe, even if the sensor is permitted to apply an arbitrary pair of probe fields within its volume. This limitation occurs due to probe field interference, as before for the boundary probe in Sec. \ref{sec:probeinter}. That is, including more than three pairs of boundary modes breaks the isotropy of $\psi(\boldsymbol{r})$ in the exterior, and a sensor can always improve upon an anisotropic $\psi(\boldsymbol{r})$ by performing multiple rotated copies of the probe and combining the results using Eq. (\ref{eq:psicoefs1}). Moreover, numerical optimization suggests that Eq. (\ref{eq:thefundamentalbound}) does not provide a close lower bound on the precision of a single volume probe, even when the continuity constraints on $V_f(\boldsymbol{r})$ and $V_w(\boldsymbol{r})$ are relaxed to allow the sensor to separately optimize $\psi(\boldsymbol{r})$ in the interior and the exterior (see Fig. 3(a) and Supplemental Material Sec. \ref{sec:bulknum}). Thus, a sensor that can perform sensory multiplexing appears to have a substantial advantage over a standard sensor.

\section{The precision of biomechanical sensing} \label{sec:mechanosensing}

In this section, we apply our modeling framework to a scenario in which structural heterogeneity is known to play a significant role: cellular mechanosensing. Certain types of eukaryotic cells engage in mechanosensing by actively probing and responding to the stiffness of their surroundings \cite{Discher2005,Vogel2006a}. These mechanical cues have been shown to govern cellular behavior in decisive ways, including guiding cell migration \cite{Lo2000, Isenberg2009} and determining cell fate \cite{Engler,Guilak2009}. However, it is an open question whether cells have evolved to make optimal use of the mechanical information available to them. This optimality hypothesis has led to strikingly successful predictions of cellular behavior in the context of chemical sensing \cite{Berg1977,Endres2008,Bialek2005,Petkova2019}.

In connective tissue, a cell's mechanical environment primarily consists of a disordered biopolymer network that serves as a scaffold on which the cell lives and moves \cite{Frantz2010}. Although the local response of such networks has been well-characterized in both experiment and theory \cite{Head2005, Doyle2016,Beroz2017}, the extent to which these local cues allow cells to infer global mechanical properties has remained unclear.

 To quantify what a cell can learn by interacting with a biopolymer network, we consider our sensing model for a three-dimensional, isotropic elastic medium characterized by a shear modulus $\mu$ and a Poisson's ratio $\sigma$. For simplicity, we take $\sigma$ to be a fixed, uniform field and $\mu$ to the sum of a fixed, uniform field $\mu_0$ and a spatially-varying Gaussian random field $\delta\mu(\boldsymbol{r})$ with spatial correlations given by:

\begin{equation}\label{eq:dldl}
\langle \delta \mu(\boldsymbol{r}) \delta \mu(\boldsymbol{r}') \rangle = \Delta_{\mu}\xi^D \delta (\boldsymbol{r} - \boldsymbol{r}').
\end{equation}

\noindent The internal energy of such an elastic solid is given by:

\begin{dmath} \label{eq:intelsol}
E = \int \mu(\boldsymbol{r}) \left( \frac{1}{2} \partial_i u_k(\boldsymbol{r}) \partial_i u_{k}(\boldsymbol{r}) +  \frac{1}{2} \partial_i u_k(\boldsymbol{r}) \partial_k u_{i}(\boldsymbol{r}) + \frac{\varsigma}{2}  \partial_i u_{i}(\boldsymbol{r})\partial_k u_{k}(\boldsymbol{r}) \right) d\boldsymbol{r},
\end{dmath}

\noindent where $u_i(\boldsymbol{r})$ is the deformation vector field, $\varsigma = 2\sigma / (1 - 2\sigma)$ is a constant, and repeated indices imply summation from $1$ to $3$ over the indexed terms.

Eukaryotic cells attach to biopolymer networks via transmembrane protein complexes called focal adhesions, which allow the cell to sense stiffness \cite{Vogel2006a,Fletcher2010,Trichet2012,Doyle2016}. We model the cell as an idealized stiffness-measuring device that first applies a force vector field $f_i(\boldsymbol{r})$:

\begin{equation}\label{eq:fspring2}
\delta E = - \int f_i(\boldsymbol{r}) u_i(\boldsymbol{r}) d\boldsymbol{r}.
\end{equation}

\noindent Once the medium reaches mechanical equilibrium, we assume that the cell transduces the following integrated response $m$:

\begin{equation}\label{eq:mspringcont}
m = \int w_i(\boldsymbol{r}) u_i(\boldsymbol{r}) d\boldsymbol{r},
\end{equation}

\noindent where $w_i(\boldsymbol{r})$ is a weight vector field. In what follows, we will estimate the precision with which a cell can infer $\mu_0$ based on $m$ and prior knowledge of all other model parameters (including $\sigma$). For this sensory process, the optimal estimator is given by Eq. (\ref{eq:haha}) with the following probe intensity:

\begin{dmath}
\psi(\boldsymbol{r}) = \partial_i V_{f,k}(\boldsymbol{r}) \partial_i V_{w,k}(\boldsymbol{r}) +  \partial_i V_{f,k}(\boldsymbol{r}) \partial_k V_{w,i}(\boldsymbol{r}) + \varsigma \partial_i V_{f,i}(\boldsymbol{r}) \partial_k V_{w,k}(\boldsymbol{r}),
\end{dmath}

\noindent where $V_{f,i}(\boldsymbol{r})$ and $V_{w,i}(\boldsymbol{r})$ are the probe potentials that correspond to the probe vector fields $f_i(\boldsymbol{r})$ and $w_i(\boldsymbol{r})$, respectively (see Supplemental Material, Sec. \ref{sec:pice}).

In mechanical equilibrium, a cell cannot exert a net force on the medium due to the requirement of force balance. Under this restriction, a cell in $D=3$ maximizes its effective range by applying probe potentials that decay as $\sim 1/r^2$ in the far-field limit. The simplest possible measurement protocol with such a profile consists of isotropic dipolar shells of radius $a$:

\begin{equation}
f_i(\boldsymbol{r}) \sim \delta(r-a) \hat{\boldsymbol{r}},
\end{equation}

\begin{equation}
w_i(\boldsymbol{r}) \sim \delta(r-a) \hat{\boldsymbol{r}}.
\end{equation}

\noindent These probe vector fields produce the following probe intensity:

\begin{equation} \label{eq:psiint7}
\psi(\boldsymbol{r})  \sim \begin{cases}
1 + 3\varsigma/2, & \text{$r<a$}.\\
2 r^{-6}, & \text{$r>a$}.
\end{cases}
\end{equation}

\noindent We insert this probe intensity into Eq. (\ref{eq:varspring}) to obtain the following variance $\delta\mu_0^2$ in the cell's estimate of $\mu_0$:

\begin{equation} \label{eq:cellvar}
\delta\mu_0^2 = \Delta_{\mu}\xi^D V^{-1} \left(\frac{27\varsigma^2 + 36 \varsigma + 28}{27\varsigma^2 + 108\varsigma + 108} \right).
\end{equation}

\noindent We determined the values of the parameters in our model for a reconstituted collagen network, an \emph{in vitro} system that closely resembles \emph{in vivo} cellular environments \cite{Doyle2016,Zaman2006,Guo2013,Beroz2017}. For a collagen network prepared from a $c \sim \SI{0.2}{\microgram/\milli \liter}$ solution of collagen type-I monomers, previous studies suggest $\mu_0\sim \SI{0.3}{\pascal}$, $\sigma \sim 0.4$, $\Delta_{\mu} \sim \SI{0.1}{\pascal^2}$, and $\xi \sim \SI{5}{\micrometer}$ (see Supplemental Material, Sec. \ref{sec:params}). For these values, the ratio $\Delta_\mu^{1/2}/\mu_0 \sim 1$ lies outside the regime of validity of our perturbative approach; nevertheless, we expect Eq. (\ref{eq:cellvar}) to provide a qualitative description of how $\delta \mu_0 / \mu_0$ depends on the model parameters.

Taking the cell radius to be $a = \SI{10}{\micrometer}$ in Eq. (\ref{eq:cellvar}) leads to a fractional uncertainty $\delta \mu_0 / \mu_0 \sim 0.15$. Thus, our theoretical framework supports the notion that cells could use mechanical information to reliably distinguish between different connective tissue environments, including brain ($\mu_0 \sim \SI{1}{\kilo\pascal}$), muscle ($\mu_0 \sim \SI{10}{\kilo\pascal}$), and bone ($\mu_0 \sim \SI{100}{\kilo\pascal}$) \cite{Lo2000,Engler,Isenberg2009,Guilak2009,Doyle2016}. Such mechanosensing could be tested in experiment by using micropatterned materials to explore the effect of substrate heterogeneity on intracellular signaling dynamics, e.g. as in Ref. \cite{Yang2016}.

In principle, a cell could reduce its measurement uncertainty via sensory multiplexing. To that end, the spatial resolution of the cell is limited by the maximum number of focal adhesions that it can simultaneously apply to the network. Interestingly, cells have been observed to display more than $\sim 100$ focal adhesions \cite{Prager-Khoutorsky2011}, which could allow a $\SI{10}{\micrometer}$ cell to probe the network on scales smaller than $\xi$. As our analysis only applies for $d \gg \xi$, we take $d \sim \SI{5}{\micrometer}$ to estimate the improvement in the cell's precision. Using this value, the reasoning in Sec. \ref{sec:smsolid} of the Supplemental Material suggests that sensory multiplexing could improve the fractional uncertainty of the cell's estimate by roughly a factor of three, down to $\delta \mu_0 / \mu_0 \sim 0.05$. This value is comparable to the smallest relative differences in bulk stiffness that elicit significant changes in cellular motility and differentiation on homogeneous substrates \cite{Hadden2017}. Thus, our results suggest that adapting such behavioral assays to heterogeneous substrates could provide an experimental test of whether cells employ sensory multiplexing.

\section{Discussion}

Measurement is a cornerstone of science. However, the physical significance of a measurement cannot be assessed without a characterization of its uncertainty. In light of this fact, even the most elementary observations of material properties are ambiguous, as materials always possess some degree of heterogeneity. This philosophical issue has rapidly evolved into a practical issue as advances in technology have brought the microscopic realm to the forefront of industry and science \cite{Fahlbuscha,Maex2003,Cullinan2012,JingW2014Real, Wallace2017,Bhat2004,Nelson2006,Ornes2017,Simaan2018,Discher2005,Arlett2011,Skedung2013,Yang2016,Beroz2017,Petridou2017}. In this regime, measurements can no longer be assumed to self-average over material heterogeneities. The effect of these heterogeneities on the ability to perceive material properties has remained unclear.

To address this gap in our understanding, we have developed a theoretical framework for calculating what a physical sensor can learn by interacting with a material. By applying our framework to several examples of random media, we have elucidated how a medium's response function can govern the limits of a sensor's precision. In particular, we found that the fractional uncertainty of a sensor's estimate of a material constant $\lambda_0$ is bounded by Eq. (\ref{eq:fuwinkler}) for a short-ranged response function and Eq. (\ref{eq:funcert}) for a long-ranged response function. Remarkably, Eq. (\ref{eq:funcert}) implies that a finite-sized sensor can achieve arbitrarily high precision --- in effect, averaging the material constant field $\lambda(\boldsymbol{r})$ over an arbitrarily large volume --- provided that it can probe the medium on small enough scales $d$. To reach this bound, the sensor must execute multiple, distinct measurements and make an estimate of $\lambda_0$ based on their combined outcomes using Eqs. (\ref{eq:psicoefs1}) and (\ref{eq:psicoefs2}). This ``sensory multiplexing'' provides a novel design principle for engineering high precision sensors that would be well-suited for applications on the microscopic scale \cite{Fahlbuscha,Maex2003,Cullinan2012,JingW2014Real, Wallace2017,Bhat2004,Nelson2006,Ornes2017,Simaan2018}. Finally, we have applied our framework to an \emph{in vitro} model for connective tissue to elucidate the limits of cellular mechanosensing, a sensory process that is known to guide cellular differentiation and motility \cite{Lo2000,Engler,Isenberg2009,Guilak2009,Doyle2016}.

Our framework only relies on a few basic assumptions, and so we expect our results to be relevant for sensory processes in a wide variety of media with quenched, random disorder. For simplicity, we focused on spherical sensors embedded inside a medium at a fixed location. However, our framework can also be used to study different sensory geometries and motile sensors. Many sensors operate on the boundary of media, including cells grown on flat surfaces \cite{Yang2016}. Moreover, cells in connective tissue can become highly elongated \cite{Prager-Khoutorsky2011} and undergo directed migration \cite{Wu2014}, both of which may serve as behavioral strategies for overcoming spatial correlations in the material properties.

To gain analytical insight into sensing, we made specific assumptions about the media we considered. Throughout the main text, we assumed a sufficiently small material correlation length $\xi$, which holds provided that $d \gg \xi$. Our approach can be readily extended to account for a finite correlation length $\xi$, which we have done for a Winkler foundation in Supplemental Material Sec. \ref{sec:winklercor}. Moreover, although we have focused mostly on a simple scalar version of elasticity, we expect our scaling results to hold for a broad range of media with long-ranged response functions, including the three-dimensional elastic medium in Sec. \ref{sec:mechanosensing}. Finally, we assumed that the elastic properties of the medium within the sensing volume are not significantly mismatched from those of the exterior. Extending our model to account for more complicated constitutive relations and other distributions of disorder are important directions for future research.

Our theory can be used to better quantify the human capacity for sensing by touch. Previous psychophysical experiments have found that human sensory systems operate at or near the physical limits of resolution \cite{Skedung2013,Tinsley2016}. For example, our ability to distinguish topographical features via tactile sensation extends down to the nanoscale \cite{Tinsley2016}. It would be interesting to determine the extent to which the human brain is capable of interpreting mechanical cues to perform sensory multiplexing.

We have focused on athermal materials. For thermal materials, the quantities measured by the sensor fluctuate in time. These fluctuations provide an additional source of temporal noise to the inference process, as well as additional response configurations that can be observed by the sensor. Generalizing our approach to account for these effects would provide a comprehensive physical limit to sensing the properties of materials.

In summary, we have elucidated the perception of material properties in physical space. On small scales, structural heterogeneities place limits on the precision of sensing. Going forward, our theory will guide the design of the next generation of sensors that will be capable of probing materials at the fundamental limits of spatial resolution.

%\newpage
%\onecolumngrid
%\newpage

\begin{acknowledgments}
	We thank Aris Alexandradinata, Masud Beroz, William Bialek, Chase Broedersz, Judith H\"oller, David Huse, Tim (Hou Keong) Lou, Yigal Meir, Joshua Shaevitz, Ian Tobasco, and Ned Wingreen for insightful comments and discussions. This work was supported in part by the National Science Foundation Grants DMR-1056456 (to D.K.L.), DMR 1609051 (to X.M.), and EFRI-1741618 (to D.Z. and X.M.), a Margaret and Herman Sokol Faculty Award (to D.K.L.), and a Michigan Life Sciences fellowship (to F.B.).
\end{acknowledgments}

\bibliography{true-limits}

\newpage

\newpage

\title{Physical limits to sensing material properties - Supplemental Material}

\author{Farzan Beroz}
\email{Corresponding author. Email: farzan@umich.edu}
\affiliation{Department of Physics, University of Michigan, Ann Arbor, Michigan 48109, USA}
\author{Di Zhou}
\affiliation{Department of Physics, University of Michigan, Ann Arbor, Michigan 48109, USA}
\author{Xiaoming Mao}
\affiliation{Department of Physics, University of Michigan, Ann Arbor, Michigan 48109, USA}
\author{David K. Lubensky}
\affiliation{Department of Physics, University of Michigan, Ann Arbor, Michigan 48109, USA}

\maketitle

\renewcommand{\figurename}{Supplementary Figure}
\renewcommand{\theequation}{S\arabic{equation}}

\titleformat{\section}    
{\normalfont\fontfamily{cmr}\fontsize{12}{17}\bfseries}{\thesection}{1em}{}
\titleformat{\subsection}[runin]
{\normalfont\fontfamily{cmr}\bfseries}{}{1em}{}

\renewcommand\thefigure{\arabic{figure}}    
\renewcommand\thesection{S\arabic{section}}

\setcounter{section}{0}
\setcounter{figure}{0}
\setcounter{equation}{0}

\setlength{\parskip}{0pt}

\newpage

\onecolumngrid

\newpage

\pagebreak

\onecolumngrid
\begin{center}
	\textbf{\large Physical limits to sensing material properties - Supplemental Material}\\[.2cm]
	Farzan Beroz,$^{1,*}$ Di Zhou,$^{1}$ Xiaoming Mao,$^{1}$ and David K. Lubensky$^1$\\[.1cm]
	{\itshape ${}^1$Department of Physics, University of Michigan, Ann Arbor, Michigan 48109\\
		}
	${}^*$Corresponding author. Email: farzan@umich.edu
\end{center}

\section{The optimal estimator for the Winkler foundation} \label{sec:optwink}

In this section, we prove that $\hat{\lambda}_0 = s/m$ is the optimal, minimum-variance unbiased estimator (MVUE) of $\lambda_0$ for the Winkler foundation. We assume that the sensor has prior knowledge of the model parameters $a$, $\xi$, and $\Delta_{\lambda}$, as well as of the configurations of the probe fields $f(\boldsymbol{r})$ and $w(\boldsymbol{r})$. We insert Eqs. (\ref{eq:mspring}) and (\ref{eq:sspring}) into Eq. (\ref{eq:haha}) to find:

\begin{equation} \label{eq:appenda}
\hat{\lambda}_0 = \frac{\int w(\boldsymbol{r}) f(\boldsymbol{r}) d\boldsymbol{r}}{\int w(\boldsymbol{r}) u(\boldsymbol{r}) d\boldsymbol{r}}.
\end{equation}

\noindent Here, $u(\boldsymbol{r})$ is the response field of the medium in mechanical equilibrium, i.e the solution of:

\begin{equation}
\frac{\delta}{\delta u} (E + \delta E) = \lambda(\boldsymbol{r}) u(\boldsymbol{r}) - f(\boldsymbol{r}) = 0.
\end{equation}

\noindent Thus, $u(\boldsymbol{r}) = f(\boldsymbol{r}) / \lambda(\boldsymbol{r})$. We take the Taylor expansion of this response field to find:

\begin{equation}
u(\boldsymbol{r}) = \frac{f(\boldsymbol{r})}{\lambda_0} \left(1-\frac{\delta\lambda(\boldsymbol{r})}{\lambda_0} \right),
\end{equation}

\noindent to leading order in $\delta \lambda(\boldsymbol{r})$. Inserting this response field into Eq. (\ref{eq:appenda}) and performing another Taylor expansion yields:

\begin{equation} \label{eq:appest}
\hat{\lambda}_0 = \lambda_0 + \frac{ \int \delta\lambda(\boldsymbol{r}) \psi(\boldsymbol{r}) d\boldsymbol{r} }{\int \psi(\boldsymbol{r}) d\boldsymbol{r}   }  ,
\end{equation}

\noindent to leading order in $\delta \lambda(\boldsymbol{r})$, where $\psi(\boldsymbol{r}) = f(\boldsymbol{r})w(\boldsymbol{r})$. To prove that $\hat{\lambda}_0$ is the MVUE for $\lambda_0$, we invoke the Lehmann-Scheff\'e theorem, which states that $\hat{\lambda}_0$ is the MVUE for $\lambda_0$ if $\hat{\lambda}_0$ is an (\emph{i}) unbiased, (\emph{ii}) sufficient, and (\emph{iii}) complete statistic for $\lambda_0$ \cite{Lehmann2011,Lehmann2011a}. We consider these conditions in turn:

\subsection*{(\emph{i}) $\hat{\lambda}_0$ is unbiased.}
\ \\

An unbiased estimator is equal to the estimated quantity on average \cite{Younga}:

\begin{equation}
\langle \hat{\lambda}_0 \rangle = \lambda_0.
\end{equation}

\noindent This equivalence can be shown starting from Eq. (\ref{eq:appest}) as follows:

\begin{equation}
\langle \hat{\lambda}_0 \rangle = \lambda_0 + \frac{ \int  \langle \delta\lambda(\boldsymbol{r}) \rangle \psi(\boldsymbol{r}) d\boldsymbol{r} }{\int \psi(\boldsymbol{r}) d\boldsymbol{r}   }  = \lambda_0,
\end{equation}

\noindent where the second equality follows from our definition of $\delta \lambda(\boldsymbol{r})$ as a Gaussian random field with zero mean.

\subsection*{(\emph{ii}) $\hat{\lambda}_0$ is sufficient.}
\ \\

The sufficiency of $\hat{\lambda}_0$ can be established using the Fisher factorization theorem \cite{Fisher1922}. According to this theorem, a function $\hat{\lambda}_0(\boldsymbol{X})$ of the observed data $\boldsymbol{X}$ is a sufficient statistic for $\lambda_0$ if and only if:

\begin{equation}\label{eq:fisher}
P(\boldsymbol{X} | \lambda_0) = u(\boldsymbol{X}) v(\hat{\lambda}_0(\boldsymbol{X}), \lambda_0),
\end{equation}

\noindent where $u$ is a nonnegative function that depends only on the data $\boldsymbol{X}$ and $v$ is a nonnegative function that can depend on $\boldsymbol{X}$ as well as $\lambda_0$, but for which the only dependence on $\boldsymbol{X}$ is through $\hat{\lambda}_0$.

Here, we take $\boldsymbol{X} = 1/m$. The probability distribution for $\boldsymbol{X}$ conditional on the value of $\lambda_0$ is given by:

\begin{equation}
P(\boldsymbol{X} | \lambda_0) = \frac{1}{\sqrt{2\pi \delta \lambda_0^2 / s^2}  } e^{-\frac{(\boldsymbol{X} - \lambda_0/s)^2}{2\delta \lambda_0^2 / s^2}   }.
\end{equation}

\noindent We factor this expression as follows:

\begin{equation}
P(\boldsymbol{X} | \lambda_0) = \left( \frac{1}{\sqrt{2\pi \delta \lambda_0^2 / s^2}  } e^{-\frac{\boldsymbol{X}^2}{2\delta \lambda_0^2 / s^2} } \right)  \left( e^{ \frac{s \boldsymbol{X} \lambda_0 }{\delta \lambda_0^2} - \frac{\lambda_0^2}{2\delta \lambda_0^2 }  }\right).
\end{equation}

\noindent We now substitute the estimator $\hat{\lambda}_0 = s\boldsymbol{X}$ into the second exponent to find:

\begin{equation}
P(\boldsymbol{X} | \lambda_0) = \left( \frac{1}{\sqrt{2\pi \delta \lambda_0^2 / s^2}  } e^{-\frac{\boldsymbol{X}^2}{2\delta \lambda_0^2 / s^2} } \right)  \left( e^{ \frac{\hat{\lambda}_0 \lambda_0 }{\delta \lambda_0^2} - \frac{\lambda_0^2}{2\delta \lambda_0^2 }  }\right).
\end{equation}

\noindent The first and second terms of this expression can be identified, respectively, with the functions $u$ and $v$ in Eq. (\ref{eq:fisher}), which demonstrates the condition (\emph{ii}) of sufficiency.

\subsection*{(\emph{iii}) $\hat{\lambda}_0$ is complete.}
\ \\

A statistic has the property of completeness if the following relationship holds for every measureable function $g$ \cite{Younga}:

\begin{equation}\label{eq:completeness}
\mathrm{If}\ \langle g(\hat{\lambda}_0) | \lambda_0 \rangle = 0 \textrm{ for all }\lambda_0\textrm{,}  \textrm{then } P(g(\hat{\lambda}_0) = 0 | \lambda_0) = 1\textrm{ for all }\lambda_0,
\end{equation}

\noindent where the conditional average $\langle g(\hat{\lambda}_0) | \lambda_0 \rangle$ is given by:

\begin{equation}
\langle g(\hat{\lambda}_0) | \lambda_0 \rangle = \frac{1}{\sqrt{2\pi \delta \lambda_0^2}} \int_{-\infty}^{\infty} g(\hat{\lambda}_0) e^{-\frac{(\hat{\lambda}_0 - \lambda_0)^2}{2\delta \lambda_0^2}}  d\hat{\lambda}_0.
\end{equation}

\noindent We factor this expression to obtain:

\begin{equation} \label{eq:twosided1}
\langle g(\hat{\lambda}_0) | \lambda_0 \rangle = k(\lambda_0) \int_{-\infty}^{\infty} h(\hat{\lambda}_0) e^{\frac{\hat{\lambda}_0  \lambda_0}{\delta \lambda_0^2}}  d\hat{\lambda}_0,
\end{equation}

\noindent where we have defined the functions:

\begin{equation}
k(\lambda_0) = \frac{1}{\sqrt{2\pi \delta \lambda_0^2}} e^{-\frac{\lambda_0^2}{2\delta \lambda_0^2}},
\end{equation}

\noindent and

\begin{equation}
h(\hat{\lambda}_0) = g(\hat{\lambda}_0) e^{-\frac{\hat{\lambda}_0^2}{2\delta \lambda_0^2}}.
\end{equation}

\noindent From Eq. (\ref{eq:twosided1}), we see that $\langle g(\hat{\lambda}_0) | \lambda_0 \rangle $ is proportional to the two-sided Laplace transform $\mathcal{L}\{ h(\hat{\lambda}_0) \} (\lambda_0 / \delta \lambda_0^2)$ of $h(\hat{\lambda}_0)$ \cite{Bellman1984}:

\begin{equation}
\langle g(\hat{\lambda}_0) | \lambda_0 \rangle = k(\lambda_0) \mathcal{L}\{ h(\hat{\lambda}_0) \}(\lambda_0 / \delta \lambda_0^2).
\end{equation}

\noindent To determine whether Eq. (\ref{eq:completeness}) is satisfied, we set this conditional average equal to zero:

\begin{equation}
0 = k(\lambda_0) \mathcal{L}\{ h(\hat{\lambda}_0) \}(\lambda_0 / \delta \lambda_0^2).
\end{equation}

\noindent The prefactor $k(\lambda_0)$ is always positive and therefore we can divide both sides by it to find:

\begin{equation}
0 = \mathcal{L}\{ h(\hat{\lambda}_0) \}(\lambda_0 / \delta \lambda_0^2).
\end{equation}

\noindent The two-sided Laplace transform is one-to-one \cite{Chareka2007}, and so we must have $h(\hat{\lambda}_0) = 0$ for all values of $\lambda_0$. However, since $h(\hat{\lambda}_0)$ is given by $g(\hat{\lambda}_0)$ times a function that is positive for all values of $\lambda_0$, we must have $g(\hat{\lambda}_0) = 0$ for all values of $\lambda_0$. Thus, the relation Eq. (\ref{eq:completeness}) is satisfied, and $\hat{\lambda}_0$ is a complete statistic for $\lambda_0$.

Taken together, the (\emph{i}) unbiasedness, (\emph{ii}) sufficiency, and (\emph{iii}) completeness of $\hat{\lambda}_0$ imply that it is the MVUE of $\lambda_0$.

\section{The probe intensity for the elastic sheet} \label{sec:pies}

The elastic sheet we considered in the main text obeys the following constitutive relation in mechanical equilibrium:

\begin{equation}
\nabla \cdot (\lambda(\boldsymbol{r}) \nabla u(\boldsymbol{r})) = f(\boldsymbol{r}).
\end{equation}

\noindent We invert this constitutive relation and perform a Taylor expansion to obtain the response field:

\begin{equation}
u(\boldsymbol{r}) =  \frac{1}{\lambda_0}V_f(\boldsymbol{r})  - \frac{1}{\lambda_0^2} \int G(\boldsymbol{r} - \boldsymbol{r'} ) \nabla' \cdot ( \delta \lambda (\boldsymbol{r'}) \nabla' V_f(\boldsymbol{r'})  ) d\boldsymbol{r'} ,
\end{equation}

\noindent to leading order in $\delta\lambda(\boldsymbol{r})$. We take the product of this response field and the weight field and integrate to find the integrated measurement $m$:

\begin{equation}
m = \int \biggl( \frac{w(\boldsymbol{r}) V_f(\boldsymbol{r})}{\lambda_0}  - \frac{1}{\lambda_0^2} \int w(\boldsymbol{r})G(\boldsymbol{r} - \boldsymbol{r'} ) \nabla' \cdot ( \delta \lambda (\boldsymbol{r'}) \nabla' V_f(\boldsymbol{r'})  ) d\boldsymbol{r'}  \biggr) d\boldsymbol{r}.
\end{equation}

\noindent In terms of the weight potential, the above expression becomes:

\begin{equation}
m = \int \left( \frac{\nabla^2V_w(\boldsymbol{r}) V_f(\boldsymbol{r}) }{\lambda_0} - \frac{1}{\lambda_0^2} V_w(\boldsymbol{r}) \nabla \cdot ( \delta \lambda (\boldsymbol{r}) \nabla V_f(\boldsymbol{r}) )    \right) d\boldsymbol{r}.
\end{equation}

\noindent To cast this expression into a form analogous to $m$ for the Winkler foundation in Eq. (\ref{eq:mspringint}), we integrate both terms by parts and find:

\begin{equation}
m = -\int \left( \frac{1 }{\lambda_0} - \frac{\delta \lambda (\boldsymbol{r})}{\lambda_0^2}   \right)  \nabla V_w(\boldsymbol{r}) \cdot \nabla V_f(\boldsymbol{r})d\boldsymbol{r}.
\end{equation}

\noindent This expression does not contain boundary terms because we have stipulated that the probe potentials must both vanish at infinity. From this expression, we identify the probe intensity as $\psi(\boldsymbol{r}) = \nabla V_w(\boldsymbol{r}) \cdot \nabla V_f(\boldsymbol{r})$. For simplicity, we have not included the minus sign in the definition of $\psi(\boldsymbol{r})$, as a change in sign does not impact the estimator $\hat{\lambda}_0$. By analogy to the Winkler foundation, it follows that the sensor can obtain an unbiased estimate of ${\lambda}_0$ for the elastic sheet by inserting this probe intensity into Eq. (\ref{eq:haha}).

\section{Numerical minimization of $\delta\lambda_0 / \lambda_0$ for boundary probes} \label{sec:snum}

To determine the optimal measurement protocol for a sensor that can apply arbitrary probe fields on its boundary, we used Mathematica's \texttt{NMinimize} function to search for a global minimum of Eq. (\ref{eq:tensorone}) subject to the constraint Eq. (\ref{eq:tensortwo}). We performed this minimization over the coefficients $B_k^{(f)}$ and $B_k^{(w)}$ using the built-in Nelder-Mead method. The accuracy and precision goals were both chosen to be $\epsilon = 8$, and we took the maximum number of iterations to be $N_{\mathrm{max}} = 1000$. To explore different local minima, we introduced stochasticity by repeating the minimization for $25$ random initial seeds for each choice of the parameter $k_{\mathrm{max}}$ defined in the main text. For each value of $k_{\mathrm{max}}$, the minimum fractional uncertainty $\delta \lambda_{0,\mathrm{min}}/\lambda_0$ reported in the main text was taken to be the minimum of the values found among the $25$ trials. We have made the Mathematica notebook used to minimize $\delta \lambda_0 / \lambda_0$ available freely on GitHub (https://github.com/farzanb/sensing-in-random-media).

\section{Measurement protocols containing more than three mode pairs break isotropy} \label{sec:breakisotropy}

In the main text, we showed that incorporating discordant modes into a boundary probe generically yields configurations of $\psi(\boldsymbol{r})$ that vary as a function of the angular coordinate (cf. Eq. (\ref{eq:beats})). Here, we demonstrate that although it is possible to cast an isotropic $\psi(\boldsymbol{r})$ using two discordant mode pairs, three or more discordant mode pairs must necessarily break isotropy. To prove this statement, we map the requirement of isotropy onto a constraint satisfaction problem. For this purpose, it is useful to begin by representing the probe potentials in terms of trigonometric functions. To be concrete, we consider the effect of boundary modes in the exterior. Here, the most general probe potentials can be represented by:

\begin{equation}
V_f(\boldsymbol{r}) = \sum_{k=1}^{k_{\mathrm{max}}} B^{(f)}_k \left(\frac{a}{r}\right)^k \cos (k \theta + \phi_k^{(f)}) ,
\end{equation}

\begin{equation}
V_w(\boldsymbol{r}) = \sum_{k=1}^{k_{\mathrm{max}}} B^{(w)}_k \left(\frac{a}{r}\right)^k \cos (k \theta + \phi_k^{(w)}).
\end{equation}

\noindent In this representation, the coefficients $B^{(f)}_k$ and $B^{(w)}_k$ are taken to be real. Finally, we take each coefficient to be positive. This choice can be made without loss of generality because each term in these expansions is invariant with respect to a change in sign combined with a phase shift of $\pi$. These probe potentials yield a probe intensity given by:

\begin{multline}
	\psi(\boldsymbol{r}) = \sum_{k,l} B^{(f)}_k B^{(w)}_k k^2 \left( \frac{a}{r}\right)^{2|k| + 2} \cos( \phi_k^{(f)} -  \phi_k^{(w)}) + B^{(f)}_l B^{(w)}_l l^2 \left( \frac{a}{r}\right)^{2|l|+ 2 } \cos ( \phi_l^{(f)}- \phi_l^{(w)} ) \\
	+ B^{(f)}_k B^{(w)}_l
	|k| |l| \left( \frac{a}{r}\right)^{|k| + |l| + 2} \cos ((k-l)\theta  + \phi_k^{(f)}  -  \phi_l^{(w)} )+  
	B^{(f)}_l B^{(w)}_k |k| |l|  \left( \frac{a}{r}\right)^{|k| + |l| + 2 } \cos ( (l-k)\theta +  \phi_l^{(f)} -  \phi_k^{(w)}).
\end{multline}

\noindent From this expression, we see that in order for $\psi(\boldsymbol{r})$ to be independent of the angular coordinate $\theta$, there must be a complete cancellation among the terms arising from the second line. To achieve such a cancellation, the following equation must be satisfied:

\begin{equation} \label{eq:consat1}
B^{(f)}_l B^{(w)}_k e^{i (\phi_l^{(f)} -  \phi_k^{(w)})} + B^{(f)}_k B^{(w)}_l  e^{-i (\phi_k^{(f)} -  \phi_l^{(w)})}  = 0,
\end{equation}

\noindent for all $k \neq l$. In the special case where the probe potentials contain only two mode pairs, these equations can be satisfied if the phases obey:

\begin{equation}
\phi_k^{(f)} + \phi_l^{(f)} =  \phi_k^{(w)} + \phi_l^{(w)} + \pi,
\end{equation}

\noindent along with the appropriate choice $B^{(f)}_k B^{(w)}_l = B^{(f)}_l B^{(w)}_k $ of amplitudes. These choices result in the following probe intensity:

\begin{equation}
\psi(\boldsymbol{r}) = \frac{B^{(f)}_l}{B^{(w)}_l} \left( B^{(w)2}_l l^2 \left( \frac{r}{a} \right)^{2k} - B^{(w)2}_k k^2 \left( \frac{r}{a} \right)^{2l}  \right) \left( \frac{a}{r} \right)^{2 + 2k + 2l} \cos ( \phi_2^{(f)} -  \phi_2^{(w)}),
\end{equation}

\noindent which is isotropic. Interestingly, however, for three or more mode pairs, Eq. (\ref{eq:consat1}) cannot be simultaneously satisfied without taking:

\begin{equation}
\phi_k^{(f)} =  \phi_k^{(w)} + \frac{\pi}{2},
\end{equation}

\noindent for each mode $k$. This choice yields $\psi(\boldsymbol{r}) = 0$. Thus, incorporating three or more nonzero mode pairs into a measurement protocol cannot yield a nonzero $\psi(\boldsymbol{r})$ with radial symmetry. This geometrical frustration among modes is analogous to geometrical frustration among spins, which precludes arrangements of three or more spins for which each spin is antiparallel with every other spin.

\section{The convex relaxation of $\delta\lambda_0 / \lambda_0$ for boundary probes} \label{sec:convexrelax}

The fractional uncertainty $\delta \lambda_0 / \lambda_0$ is a nonconvex function of the coefficients $B^{(f)}_k$ and $B^{(w)}_k$ in Eq. (\ref{eq:tensorone}) and its constraint Eq. (\ref{eq:tensortwo}). In this section, we derive an equation for the optimal configurations of $\psi(\boldsymbol{r})$ for the convex relaxation of $\delta \lambda_0 / \lambda_0$ presented in the main text. For clarity, we will represent the probe fields in terms of trigonometric functions as follows:

\begin{equation} \label{eq:dphasef3}
f(\boldsymbol{r}) \sim \delta(r - a) \left( \sum_{k=1}^{k_{\mathrm{max}}}  B^{(f)}_k \cos(k\theta) +  B^{(f)}_{k+k_{\mathrm{max}}} \sin(k\theta)  \right),
\end{equation}

\begin{equation} \label{eq:dphasef4}
w(\boldsymbol{r}) \sim \delta(r - a) \left( \sum_{k=1}^{k_{\mathrm{max}}}  B^{(w)}_k \cos(k\theta) +  B^{(w)}_{k+k_{\mathrm{max}}} \sin(k\theta)  \right),
\end{equation}

\noindent as opposed to the more compact complex representations employed in the main text. Moreover, we will work in units where $a = \Delta_{\lambda} \xi^D = 1$. In this case, the probe intensity is given by:

\begin{equation} \label{eq:fracac}
\psi_{\pm}(\boldsymbol{r}) = \sum_{k,l = 1}^{2k_{\mathrm{max}}} B_{kl} \mathcal{M}_{kl}^{(\pm)},
\end{equation}

\noindent where $B_{kl} = B^{(f)}_k B^{(w)}_l$, and $\mathcal{M}^{(\pm)}$ is a $2k_{\mathrm{max}}$ by $2k_{\mathrm{max}}$ matrix given by:

\begin{align*}
	\mathcal{M}^{(\pm)}
	& = \left( \begin{matrix} \mathcal{M}^{(1,\pm)} &  \mathcal{M}^{(2,\pm)}  \\
		\mathcal{M}^{(2,\pm) \mathsf{T}  }  &  \mathcal{M}^{(1,\pm)}  \end{matrix}  \right),
\end{align*}

\noindent where

\begin{equation} \label{eq:orthog1}
\mathcal{M}^{(1,\pm)}_{kl} = k l r^{-2 \pm k \pm l} \cos((k-l)\theta),
\end{equation}

\begin{equation}  \label{eq:orthog2}
\mathcal{M}^{(2,\pm)}_{kl} = k l r^{-2 \pm k \pm l} \sin((k-l)\theta). 
\end{equation}

\noindent The positive and negative versions of these matrices correspond to the interior and the exterior, respectively. For this representation, specifying the coefficients $B^{(f)}_k$ and $B^{(w)}_k$ is equivalent to specifying an arbitrary rank-one matrix $B_{kl}$ of real coefficients. In this case, a convex relaxation equivalent to the one defined in the main text can be obtained by relaxing the rank constraint on $B_{kl}$ in Eq. (\ref{eq:fracac}) to allow it to be an arbitrary real matrix.

Upon performing the convex relaxation, we find that the top and bottom halves of $	\mathcal{M}^{(\pm)}$ provide redundant contributions to $\psi_{\pm}(\boldsymbol{r})$. Thus, we can simplify $\psi_{\pm}(\boldsymbol{r})$ by consolidating these contributions into two smaller coefficient matrices $P_{kl}$ and $Q_{kl}$ that multiply the entries in the top half of $\mathcal{M}^{(\pm)}$ as follows:

\begin{equation}
\psi_{\pm}(\boldsymbol{r}) = \sum_{k,l=1}^{k_{\mathrm{max}}} P_{kl} \mathcal{M}_{kl}^{(1,\pm)} +  Q_{kl} \mathcal{M}_{kl}^{(2,\pm)}.
\end{equation}

\noindent This expression still contains redundant entries, because $\mathcal{M}_{kl}^{(1,\pm)} = \mathcal{M}_{lk}^{(1,\pm)} $ and $\mathcal{M}_{kl}^{(2,\pm)} = -\mathcal{M}_{lk}^{(2,\pm)} $. Thus, without loss of generality, we can simplify the above expression for $\psi_{\pm}(\boldsymbol{r})$ by taking $P_{kl} = 0$ and $Q_{kl} = 0$ for $l < k$. In addition, a further simplification occurs by noting that $\mathcal{M}_{kk}^{(2,\pm)} = 0$, which implies that $\psi(\boldsymbol{r})$ and $S$ do not depend on the values of $Q_{kk}$. Thus, in what follows, we take $Q_{kk} = 0$ without loss of generality.

To determine the optimal configurations of $\psi(\boldsymbol{r})$ for this convex relaxation, we will minimize $\delta \lambda_0 / \lambda_0$ subject to the constraint $\int \psi(\boldsymbol{r}) d\boldsymbol{r} = 1$. Following Sec. \ref{sec:probewinkler}, we impose the constraint using a Lagrange multiplier $\gamma$, which yields the following action $S$:

\begin{equation} \label{eq:fracac2}
S = \frac{1}{2}  \int_{\mathcal{R}_{\mathrm{int}}} \left(\sum_{k,l = 1}^{k_{\mathrm{max}}} P_{kl} \mathcal{M}_{kl}^{(1,+)} + Q_{kl} \mathcal{M}_{kl}^{(2,+)} \right)^2 d\boldsymbol{r} + 
\frac{1}{2}  \int_{\mathcal{R}_{\mathrm{ext}}} \left(\sum_{k,l=1}^{k_{\mathrm{max}}} P_{kl} \mathcal{M}_{kl}^{(1,-)}  + Q_{kl} \mathcal{M}_{kl}^{(2,-)}\right)^2 d\boldsymbol{r} + \gamma \sum_{k=1}^{k_{\mathrm{max}}} 2\pi k P_{kk},
\end{equation}

\noindent where the first and second integrals are taken over the interior ${\mathcal{R}_{\mathrm{int}}}$ and the exterior ${\mathcal{R}_{\mathrm{ext}}}$ of the sensor, respectively, and the coefficients must satisfy:

\begin{equation} \label{eq:diffconstr}
\sum_{k=1}^{k_{\mathrm{max}}} 2\pi k P_{kk} = 1.
\end{equation}

\noindent to ensure $\int \psi(\boldsymbol{r})d\boldsymbol{r} = 1$. By the orthogonality of Eqs. (\ref{eq:orthog1}) and (\ref{eq:orthog2}), this expression simplifies to:

\begin{multline}
	S = \frac{1}{2}  \int_{\mathcal{R}_{\mathrm{int}}} \left(\sum_{k,l = 1}^{k_{\mathrm{max}}} P_{kl} \mathcal{M}_{kl}^{(1,+)} \right)^2 d\boldsymbol{r} + 
	\frac{1}{2}  \int_{\mathcal{R}_{\mathrm{int}}} \left(\sum_{k,l = 1}^{k_{\mathrm{max}}}   Q_{kl} \mathcal{M}_{kl}^{(2,+)} \right)^2 d\boldsymbol{r}\ +\ \\
	\frac{1}{2}  \int_{\mathcal{R}_{\mathrm{ext}}} \left(\sum_{k,l = 1}^{k_{\mathrm{max}}} P_{kl} \mathcal{M}_{kl}^{(1,-)}  \right)^2 d\boldsymbol{r} +
	\frac{1}{2}  \int_{\mathcal{R}_{\mathrm{ext}}} \left(\sum_{k,l = 1}^{k_{\mathrm{max}}}  Q_{kl} \mathcal{M}_{kl}^{(2,-)}\right)^2 d\boldsymbol{r} 
	+ \gamma \sum_{k=1}^{k_{\mathrm{max}}} 2\pi k P_{kk}.
\end{multline}

\noindent To minimize $S$, we start by separating the sums into diagonal and off-diagonal contributions as follows:

\begin{multline}
	S = \frac{1}{2}  \int_{\mathcal{R}_{\mathrm{int}}} \left(\sum_{k = 1}^{k_{\mathrm{max}}} P_{kk} \mathcal{M}_{kk}^{(1,+)}  + \sum_{l > k}^{k_{\mathrm{max}}} P_{kl} \mathcal{M}_{kl}^{(1,+)}  \right)^2 d\boldsymbol{r} + 
	\frac{1}{2}  \int_{\mathcal{R}_{\mathrm{int}}} \left( \sum_{l > k}^{k_{\mathrm{max}}} Q_{kl} \mathcal{M}_{kl}^{(2,+)}  \right)^2 d\boldsymbol{r}\ + \\
	\frac{1}{2}  \int_{\mathcal{R}_{\mathrm{ext}}} \left(\sum_{k = 1}^{k_{\mathrm{max}}} P_{kk} \mathcal{M}_{kk}^{(1,-)}  + \sum_{l > k}^{k_{\mathrm{max}}} P_{kl} \mathcal{M}_{kl}^{(1,-)}  \right)^2 d\boldsymbol{r} + 
	\frac{1}{2}  \int_{\mathcal{R}_{\mathrm{int}}} \left( \sum_{l > k}^{k_{\mathrm{max}}} Q_{kl} \mathcal{M}_{kl}^{(2,-)}  \right)^2 d\boldsymbol{r} + \gamma \sum_{k=1}^{k_{\mathrm{max}}} 2\pi k P_{kk}.	
\end{multline}

\noindent These diagonal and off-diagonal contributions are orthogonal to each other, which allows the following simplification:

\begin{multline}
	S = \frac{1}{2}  \int_{\mathcal{R}_{\mathrm{int}}} \left(\sum_{k = 1}^{k_{\mathrm{max}}} P_{kk} \mathcal{M}_{kk}^{(1,+)}   \right)^2 d\boldsymbol{r} + 
	\int_{\mathcal{R}_{\mathrm{int}}} \left( \sum_{l > k}^{k_{\mathrm{max}}} P_{kl} \mathcal{M}_{kl}^{(1,+)}  \right)^2 d\boldsymbol{r} + 
	\frac{1}{2}  \int_{\mathcal{R}_{\mathrm{int}}} \left( \sum_{l > k}^{k_{\mathrm{max}}} Q_{kl} \mathcal{M}_{kl}^{(2,+)}  \right)^2 d\boldsymbol{r}\ + \\
	\frac{1}{2}  \int_{\mathcal{R}_{\mathrm{ext}}} \left(\sum_{k = 1}^{k_{\mathrm{max}}} P_{kk} \mathcal{M}_{kk}^{(1,-)}  \right)^2 d\boldsymbol{r} + 
	\frac{1}{2}  \int_{\mathcal{R}_{\mathrm{ext}}} \left( \sum_{l > k}^{k_{\mathrm{max}}} P_{kl} \mathcal{M}_{kl}^{(1,-)}  \right)^2 d\boldsymbol{r} + 
	\frac{1}{2}  \int_{\mathcal{R}_{\mathrm{int}}} \left( \sum_{l > k}^{k_{\mathrm{max}}} Q_{kl} \mathcal{M}_{kl}^{(2,-)}  \right)^2 d\boldsymbol{r} + \gamma \sum_{k=1}^{k_{\mathrm{max}}} 2\pi k P_{kk}.	
\end{multline}

\noindent From this expression, we see that the off-diagonal contributions can only increase $S$. Therefore, to minimize $S$, the coefficients must satisfy $P_{kl} = 0$ and $Q_{kl} = 0$ for $k \neq l$, which correspond to isotropic configurations of $\psi(\boldsymbol{r})$. For these choices of coefficients, the above equation becomes:

\begin{equation}
S = \frac{1}{2}  \int_{\mathcal{R}_{\mathrm{int}}} \left(\sum_{k = 1}^{k_{\mathrm{max}}} P_{kk} \mathcal{M}_{kk}^{(1,+)}   \right)^2 d\boldsymbol{r} 
\frac{1}{2}  \int_{\mathcal{R}_{\mathrm{ext}}} \left(\sum_{k = 1}^{k_{\mathrm{max}}} P_{kk} \mathcal{M}_{kk}^{(1,-)}  \right)^2 d\boldsymbol{r} + \gamma \sum_{k=1}^{k_{\mathrm{max}}} 2\pi k P_{kk}.	
\end{equation}

\noindent To minimize $S$, the remaining undetermined coefficients $P_{kk}$ must satisfy $dS/dP_{kk} = 0$, which corresponds to the following equation:

\begin{equation} \label{eq:matsolve1}
\int_{\mathcal{R}_{\mathrm{int}}} \left(  \mathcal{M}^{(1,+)}_{mn} \sum_{k = 1}^{k_{\mathrm{max}}} P_{kk} \mathcal{M}_{kk}^{(1,+)} \right) d\boldsymbol{r} + \int_{\mathcal{R}_{\mathrm{ext}}} \left(  \mathcal{M}^{(1,-)}_{mn} \sum_{k=1}^{k_{\mathrm{max}}} P_{kk} \mathcal{M}_{kk}^{(1,-)} \right) d\boldsymbol{r} = -   2\pi \gamma m \delta_{m,n},
\end{equation}

\noindent where $\mathcal{M}_{kk}^{(1,\pm)}  =   k^2 r^{-2\pm2k}$. To determine the coefficients $P_{kk}$, we take $m=n$ in the above equation and perform the integrals to find:

\begin{equation}
\sum_{k=1}^{k_{\mathrm{max}}} P_{kk} C_{km} = -\gamma m,
\end{equation}

\noindent where $C_{km}$ is a matrix given by:

\begin{equation}
C_{km} = \frac{1}{2} \left(\frac{k^2 m^2 }{k + m - 1} + \frac{k^2 m^2 }{k + m + 1}   \right).
\end{equation}

\noindent Thus, the coefficients $P_{kk}$ are given by:

\begin{equation} \label{eq:bkk1}
P_{kk} = -\gamma \sum_{m=1}^{k_{\mathrm{max}}}  m C_{km}^{-1} ,
\end{equation}

\noindent where $\gamma$ can be determined by inserting the above equation into Eq. (\ref{eq:diffconstr}), which yields:

\begin{equation} \label{eq:bkk2}
\gamma = - \left( 2\pi \sum_{k,m=1}^{k_{\mathrm{max}}}  km C^{-1}_{km} \right)^{-1}.
\end{equation}

\noindent To explore the behavior of $\delta\lambda_0 / \lambda_0$ for large $k_{\mathrm{max}}$, we determined the coefficients $P_{kk}$ by numerically inverting $C_{km}$ and solving Eqs. (\ref{eq:bkk1}) and (\ref{eq:bkk2}). We then inserted the resulting $\psi(\boldsymbol{r})$ into Eq. (\ref{eq:varspring}). This calculation yields values of $\delta\lambda_0 / \lambda_0$ that rapidly converge to $\delta \lambda_{0,\mathrm{low}} / \lambda_0 \approx 1 / \sqrt{\pi}$ as $k_{\mathrm{max}}$ is increased. This convergence suggests a representation of $\pi$ that, to our knowledge, has not previously been reported in the literature. A simplified form of this representation can be obtained by rescaling the matrix $C_{km} \rightarrow \tilde{C}_{km}$, where:

\begin{equation}
\tilde{C}_{km} = \frac{k m}{k + m - 1} + \frac{k m}{k + m + 1}.
\end{equation} 

\noindent In terms of this matrix, our results suggest the following identity:

\begin{equation}
\lim_{N \rightarrow \infty} \sum_{k,m = 1}^{N} \tilde{C}_{km}^{-1} = \frac{\pi}{4}.
\end{equation}

\section{The optimal estimator for multiple probes} \label{sec:optmult}

A sensor that performs multiple probes can make a more precise estimate of $\lambda_0$. To do so, the sensor must take advantage of the correlations among the probes by adding up the results of the measurements with appropriately chosen weights, such that the deviations of different measurements from $\lambda_0$ cancel each other out. In this section, we prove that the minimum-variance unbiased estimator (MVUE) of $\lambda_0$ for a sequence of probes is given by the best linear unbiased sum of the estimators for individual probes. We then derive a representation of the MVUE in terms of an effective probe intensity and an equation for the variance of this estimator.

\subsection*{(\emph{i}) unbiased estimator of $\lambda_0$}\ \\

\noindent The most general linear estimator is given by:

\begin{equation}
\hat{\lambda}_0 = \sum_k q_k \hat{\lambda}_{0,k},
\end{equation}

\noindent where $q_k$ are constant weights assigned to each measurement protocol $k$. For $\hat{\lambda}_0$ to be unbiased, we must have $\langle \hat{\lambda}_0 \rangle = \lambda_0$, which implies:

\begin{equation}\label{eq:unbiased}
\sum_k q_k = 1.
\end{equation}

\noindent The variance $\delta \lambda_0^2$ of the estimator is given by:

\begin{equation}
\delta \lambda_0^2 = \sum_{k,l} C_{kl} q_k q_l,
\end{equation}

\noindent where $C_{kl} = \langle (\hat{\lambda}_{0,k} - \lambda_0) (\hat{\lambda}_{0,l} -  \lambda_0) \rangle$ is the covariance matrix of the estimators for the individual probes $k$. To determine the best linear estimator, we minimize this variance with respect to the weights $q_k$, subject to the constraint Eq. (\ref{eq:unbiased}). The optimal $q_k$ must satisfy the following equation:

\begin{equation}
\frac{d}{dq_k} \left( \sum_{k,l} C_{kl} q_k q_l - 2\gamma  \sum_k q_k  \right) = 0,
\end{equation}

\noindent where $2\gamma$ is a Lagrange multiplier that enforces Eq. (\ref{eq:unbiased}), and we have included the factor of $2$ for later convenience. The above equation simplifies to yield the following equation for $q_k$:

\begin{equation}
\sum_{l} C_{kl} q_l = \gamma.
\end{equation}

\noindent Thus, the optimal weights are given by:

\begin{equation}
q_k = \gamma \sum_l C^{-1}_{kl}.
\end{equation}

\noindent To solve for $\gamma$, we sum over the index $k$ and apply the unbiasedness constraint Eq. (\ref{eq:unbiased}) to find:

\begin{equation} \label{eq:gamma}
\gamma = \left( \sum_{k,l} C_{kl}^{-1} \right)^{-1}.
\end{equation}

\noindent Thus, the best unbiased linear estimator $\hat{\lambda}_0$ is given by:

\begin{equation} \label{eq:hlam}
\hat{\lambda}_0 = \frac{\sum_{k,l} C^{-1}_{kl} \hat{\lambda}_{0,k} }{\sum_{k,l} C^{-1}_{kl} }.
\end{equation}

By the Lehmann-Scheff\'e theorem \cite{Lehmann2011,Lehmann2011a}, $\hat{\lambda}_0$ is the MVUE for $\lambda_0$ if $\hat{\lambda}_0$ is an (\emph{i}) unbiased, (\emph{ii}) sufficient, and (\emph{iii}) complete statistic for $\lambda_0$. It remains to be shown that conditions (\emph{ii}) and (\emph{iii}) are satisfied.

\subsection*{(\emph{ii}) $\hat{\lambda}_0$ is sufficient.}\ \\

\noindent According to the Fisher factorization theorem \cite{Fisher1922}, $\hat{\lambda}_0(\hat{\lambda}_{0,k})$ with $\hat{\lambda}_{0,k}$ drawn from the conditional probability distribution $P(\hat{\lambda}_{0,k} | \lambda_0)$ is a sufficient statistic for $\lambda_0$ if:

\begin{equation}\label{eq:fisher2}
P(\hat{\lambda}_{0,k} | \lambda_0) = u(\hat{\lambda}_{0,k}) v(\hat{\lambda}_0(\hat{\lambda}_{0,k}), \lambda_0),
\end{equation}

\noindent where $u$ is a function that depends only on the individual estimators $\hat{\lambda}_{0,k}$ and $v$ is a function that can depend on $\hat{\lambda}_{0,k}$ as well as $\lambda_0$, but for which the only dependence on $\hat{\lambda}_{0,k}$ is through $\hat{\lambda}_0$. The probability distribution for $\hat{\lambda}_{0,k}$ is given by the following multivariate normal distribution:

\begin{equation}
P(\hat{\lambda}_{0,k} | \lambda_0) = \frac{1}{\sqrt{(2\pi)^N |C_{kl}| }} e^{-\frac{1}{2} \sum_{k,l} (\hat{\lambda}_{0,k} - \lambda_0) C_{kl}^{-1} (\hat{\lambda}_{0,k} - \lambda_0) },
\end{equation}

\noindent where $N$ is the total number of probes and $|C_{kl}|$ is the determinant of $C_{kl}$. We perform the Fisher factorization by first expanding the exponent as follows:

\begin{equation}
P(\hat{\lambda}_{0,k} | \lambda_0) = \frac{1}{\sqrt{(2\pi)^N |C_{kl}| }} e^{-\frac{1}{2} \sum_{k,l} \left(\hat{\lambda}_{0,k} C_{kl}^{-1} \hat{\lambda}_{0,l} - 2 \lambda_0 C_{kl}^{-1} \hat{\lambda}_{0,l} + \lambda_0^2 C_{kl}^{-1} \right) },
\end{equation}

\noindent We now insert Eq. (\ref{eq:gamma}) and Eq. (\ref{eq:hlam}) into the second and third terms of the exponent and factor to find:

\begin{equation}
P(\hat{\lambda}_{0,k} | \lambda_0) = \left( \frac{1}{\sqrt{(2\pi)^N |C_{kl}| }} e^{-\frac{1}{2} \sum_{k,l} \hat{\lambda}_{0,k} C_{kl}^{-1} \hat{\lambda}_{0,l}  }  \right) \left( e^{  \lambda_0 \gamma \hat{\lambda}_0 - \frac{1}{2} \lambda_0^2 \gamma } \right).
\end{equation}

\noindent The first and second terms of this expression can be identified, respectively, with the functions $u$ and $v$ in Eq. (\ref{eq:fisher2}), which demonstrates the condition (\emph{ii}) of sufficiency.

\subsection*{(\emph{iii}) $\hat{\lambda}_0$ is complete.}\ \\

\noindent A statistic has the property of completeness if the following relationship holds for every measureable function $g$ \cite{Younga}:

\begin{equation}\label{eq:completeness2}
\mathrm{If}\ \langle g(\hat{\lambda}_0) | \lambda_0 \rangle = 0 \textrm{ for all }\lambda_0\textrm{, then } P(g(\hat{\lambda}_0) | \lambda_0) = 1\textrm{ for all }\lambda_0,
\end{equation}

\noindent where the conditional average $\langle g(\hat{\lambda}_0) | \lambda_0 \rangle$ is given by:

\begin{equation}
\langle g(\hat{\lambda}_0) | \lambda_0 \rangle = \frac{1}{\sqrt{(2\pi)^N |C_{kl}|}} \int_{-\infty}^{\infty} g(\hat{\lambda}_0) e^{-\frac{1}{2} \sum_{k,l} (\hat{\lambda}_{0,k} - \lambda_0) C_{kl}^{-1} (\hat{\lambda}_{0,k} - \lambda_0) }  \prod_k d\hat{\lambda}_{0,k}.
\end{equation}

\noindent We factor this expression to obtain:

\begin{equation} \label{eq:twosided}
\langle g(\hat{\lambda}_0) | \lambda_0 \rangle = t(\lambda_0) \int_{-\infty}^{\infty} h(\{\hat{\lambda}_{0,k} \}) e^{\hat{\lambda}_0 \gamma \lambda_0}  \prod_k d\hat{\lambda}_{0,k},
\end{equation}

\noindent where we have defined the functions:

\begin{equation}
t(\lambda_0) = \frac{1}{\sqrt{(2\pi)^N |C_{kl}|}} e^{-\frac{\gamma \lambda_0^2}{2}},
\end{equation}

\noindent and

\begin{equation}
h(\{\hat{\lambda}_{0,k} \}) = g(\hat{\lambda}_0) e^{-\frac{1}{2} \sum_{k,l} \hat{\lambda}_{0,k} C_{kl}^{-1} \hat{\lambda}_{0,l} }.
\end{equation}

\noindent From Eq. (\ref{eq:twosided}), we see that $\langle g(\hat{\lambda}_0) | \lambda_0 \rangle $ is proportional to the two-sided multivariate Laplace transform $\mathcal{L}\{ h(\{\hat{\lambda}_{0,k} \}) \} (\gamma \lambda_0)$ of $h(\{\hat{\lambda}_{0,k} \})$ \cite{Bellman1984}:

\begin{equation}
\langle g(\hat{\lambda}_0) | \lambda_0 \rangle = t(\lambda_0) \mathcal{L}\{ h(\{\hat{\lambda}_{0,k} \}) \}(\gamma \lambda_0).
\end{equation}

\noindent To determine whether Eq. (\ref{eq:completeness2}) is satisfied, we now assume that this conditional average is equal to zero:

\begin{equation}
0 = t(\lambda_0) \mathcal{L}\{ h(\{\hat{\lambda}_{0,k} \}) \}(\gamma \lambda_0).
\end{equation}

\noindent The prefactor $t(\lambda_0)$ is always positive and therefore we can divide both sides by it to find:

\begin{equation}
0 = \mathcal{L}\{ h(\{\hat{\lambda}_{0,k} \}) \}(\gamma \lambda_0).
\end{equation}

\noindent The two-sided Laplace transform is one-to-one \cite{Chareka2007}, and so we must have $h(\{\hat{\lambda}_{0,k} \}) = 0$ for all values of $\lambda_0$. However, since $h(\{\hat{\lambda}_{0,k} \})$ is given by $g(\hat{\lambda}_0)$ times a function that is positive for all values of $\lambda_0$, we must have $g(\hat{\lambda}_0) = 0$ for all values of $\lambda_0$. Thus, the relation Eq. (\ref{eq:completeness2}) is satisfied, and $\hat{\lambda}_0$ is a complete statistic for $\lambda_0$.

Taken together, the (\emph{i}) unbiasedness, (\emph{ii}) sufficiency, and (\emph{iii}) completeness of $\hat{\lambda}_0$ imply that it is the MVUE for $\lambda_0$.

\subsection*{(\emph{iv}) $\hat{\lambda}_0$ in terms of $\Psi(\boldsymbol{r})$}\ \\

\noindent In the main text, we represented the optimal estimator $\hat{\lambda}_0$ for multiple probes in terms of an effective probe intensity $\Psi(\boldsymbol{r})$. Here, we demonstrate that this representation is equivalent to the representation Eq. (\ref{eq:hlam}) derived above. To do so, we insert Eq. (\ref{eq:psicoefs2}) into Eq. (\ref{eq:psicoefs1}) to find:

\begin{equation}
\hat{\lambda}_0 = \lambda_0 + \frac{\int \sum_{k,l} \left( C_{kl}^{-1} \frac{\psi_k(\boldsymbol{r})} {\int \psi_k(\boldsymbol{r}) d\boldsymbol{r}}   \right) \delta \lambda(\boldsymbol{r})  d\boldsymbol{r} }{\sum_{k,l} C_{kl}^{-1} }.
\end{equation}

\noindent We then use Eq. (\ref{eq:weightedavg}) for the estimator of a probe to simplify this expression, which yields Eq. (\ref{eq:hlam}). The coefficients $p_k$ that appear in Eq. (\ref{eq:psicoefs2}) in the main text are related to the coefficients $q_k$ above by $q_k = p_k \gamma = p_k / (\sum_{k,l} C_{kl}^{-1})$.

\subsection*{(\emph{v}) Variance of the estimator}\ \\

\noindent The variance $\delta \lambda_0^2$ of the estimator is defined by:

\begin{equation}
\delta \lambda_0^2 = \langle (\hat{\lambda}_0 - \lambda_0)^2 \rangle.
\end{equation}

\noindent We insert Eq. (\ref{eq:hlam}) into the above expression to find:

\begin{equation}
\delta \lambda_0^2 = \left\langle \left(\frac{\sum_{k_1,l_1} C^{-1}_{k_1l_1} (\hat{\lambda}_{0,k_1}-\lambda_0) }{\sum_{k_1,l_1} C^{-1}_{k_1l_1} } \right)  \left(\frac{\sum_{k_2,l_2} C^{-1}_{k_2l_2} (\hat{\lambda}_{0,k_2} - \lambda_0) }{\sum_{k_2,l_2} C^{-1}_{k_2l_2} }  \right) \right\rangle.
\end{equation}

\noindent This expression can be simplified by factoring the denominators and invoking the definition of the covariance matrix:

\begin{equation}
\delta \lambda_0^2 =  \frac{1}{ \left(\sum_{k,l} C^{-1}_{kl} \right)^2}  \sum_{k_1,l_1} \sum_{k_2,l_2} C^{-1}_{k_1l_1} C^{-1}_{k_2l_2} C_{k_1l_2} .
\end{equation}

\noindent We take the sum over the indices $k_1$ and $k_2$ to obtain:

\begin{equation}
\delta \lambda_0^2 = \frac{\sum_{k,l} C^{-1}_{kl}}{ \left(\sum_{k,l} C^{-1}_{kl} \right)^2}   ,
\end{equation}

\noindent which simplifies to Eq. (\ref{eq:cvar}) in the main text.

\section{Equivalent measurement protocol for varying a single probe field} \label{sec:eqprot}

For a set of probes $k$ with probe fields $f_k(\boldsymbol{r})$ and $w_k(\boldsymbol{r})$, the MVUE estimator of $\lambda_0$ is given by the best linear unbiased sum of the estimators $\hat{\lambda}_{0,k}$ of individual probes:

\begin{equation} \label{eq:weightcombine}
\hat{\lambda}_0 = \sum_k q_k \hat{\lambda}_{0,k},
\end{equation}

\noindent where $q_k = \sum_l C_{kl}^{-1} / \sum_{k,l} C_{kl}^{-1}$ is the weight of each individual estimator and $C_{kl} = \langle (\hat{\lambda}_{0,k} - \lambda_0) (\hat{\lambda}_{0,l} - \lambda_0 )\rangle$ is their covariance matrix. This estimator is bilinear in the probe fields, which implies that it is not possible to extract additional information by varying a single probe field $f_k(\boldsymbol{r})$ or $w_k(\boldsymbol{r})$ of an optimal measurement protocol while keeping the other fixed. This fact follows because for a fixed choice of weight field, performing a probe with the optimal stimulus field exhaustively samples the amount of information available to the sensor, and vice versa for a fixed choice of stimulus field.

To prove this statement, we consider a pair of measurement protocols that correspond to two distinct stimulus potentials $V_{f,1}(\boldsymbol{r})$ and $V_{f,2}(\boldsymbol{r})$ and a fixed weight potential $V_{w}(\boldsymbol{r})$. Combining these two measurement protocols using Eq. (\ref{eq:weightcombine}) results in an estimator with the following variance:

\begin{equation} \label{eq:eqvar1}
\delta \lambda_0^2 = \int \left( \frac{q_1^2}{s_1^2}(\nabla V_{f_1} \cdot \nabla V_w)^2 + \frac{2 q_1 q_2}{s_1 s_2}(\nabla V_{f_1} \cdot \nabla V_w)(\nabla V_{f_2} \cdot \nabla V_w) + \frac{q_2^2}{s_2^2}(\nabla V_{f_2} \cdot \nabla V_w)^2 \right) d\boldsymbol{r}.
\end{equation}

\noindent where $s_1$ and $s_2$ are the normalizing constants for each measurement protocol. On the other hand, one can always construct an equivalent measurement protocol using only a single stimulus potential $\tilde{V}_f(\boldsymbol{r})$ as follows:

\begin{equation}
\tilde{V}_f(\boldsymbol{r}) = \frac{q_1 V_{f_1}(\boldsymbol{r})}{s_1} + \frac{q_2 V_{f_2}(\boldsymbol{r})}{s_2}.
\end{equation}

\noindent This measurement protocol yields the following variance:

\begin{equation} \label{eq:eqvar2}
\delta \lambda_0^2 = \frac{\int \left( \nabla \left(\frac{q_1 V_{f_1}(\boldsymbol{r})}{s_1} + \frac{q_2 V_{f_2}(\boldsymbol{r})}{s_2} \right) \cdot \nabla V_w(\boldsymbol{r}) \right)^2 d\boldsymbol{r} }{ q_1 + q_2}.
\end{equation}

\noindent We now enforce the constraint $q_1 + q_2 = 1$ for unbiased estimates and find that the variabilities in Eqs. (\ref{eq:eqvar1}) and (\ref{eq:eqvar2}) are equal. Thus, provided that $\tilde{V}_f(\boldsymbol{r})$ and $V_{w}(\boldsymbol{r})$ minimize $\delta \lambda_0^2$, the sensor cannot further reduce $\delta \lambda_0^2$ by applying additional measurement protocols with different stimulus potentials and the same choice of $V_{w}(\boldsymbol{r})$.

\section{Sensory multiplexing for the two-dimensional elastic sheet} \label{sec:smsheet}

In this section, we determine the covariance matrices for the sensory multiplexing protocols described in the main text. For the collection of boundary probes, inserting Eq. (\ref{eq:weightedavg}) into the definition of the covariance matrix yields:

\begin{equation} \label{eq:cmn}
C_{kl} =  \Delta_{\lambda} \xi^{D} \frac{ \int \psi_k(\boldsymbol{r}) \psi_l(\boldsymbol{r}) d\boldsymbol{r}}{  \left( \int \psi_k(\boldsymbol{r}) d\boldsymbol{r} \int \psi_l(\boldsymbol{r}) d\boldsymbol{r}  \right)}.
\end{equation}

\noindent The probe fields in Eqs. (\ref{eq:multf}) and (\ref{eq:multw}) give rise to the following probe potentials:

\begin{equation}
V_{f,k}^{\mathrm{(+)}}(\boldsymbol{r}) = V_{w,k}^{\mathrm{(+)}}(\boldsymbol{r})  = \frac{1}{k} \left(\frac{r}{a}\right)^k \cos(k\theta).
\end{equation}

\begin{equation}
V_{f,k}^{\mathrm{(-)}}(\boldsymbol{r}) = V_{w,k}^{\mathrm{(-)}}(\boldsymbol{r}) =  \frac{1}{k} \left(\frac{a}{r}\right)^k \cos(k\theta).
\end{equation}

\noindent in the interior and the exterior, respectively, for $k \ge 1$. These probe potentials cast probe intensities $\psi_k(\boldsymbol{r})$ that are proportional to the diagonal terms in the matrix $\mathcal{M}_{kl}^{(\pm)}$ that appears in Sec. \ref{sec:convexrelax}. Indeed, for each value of $k_{\mathrm{max}}$, the following identification:

\begin{equation} \label{eq:pkpkkmap}
p_k \rightarrow k P_{kk},
\end{equation}

\noindent maps the effective probe intensity $\Psi(\boldsymbol{r})$ for the boundary probes in Sec. \ref{sec:greatjob} of the main text onto a probe intensity $\psi(\boldsymbol{r})$ for the convex relaxation of a single probe in Secs. \ref{sec:probeinter} and \ref{sec:convexrelax}. A similar mapping can be done for sensory multiplexing protocols that incorporate contributions from pairs of modes with unequal mode numbers, which we did not consider in the main text. Such contributions correspond to the off-diagonal elements of $\mathcal{M}_{kl}^{(\pm)}$. Thus, for a boundary probe, the possible $\Psi(\boldsymbol{r})$ that can be achieved by sensory multiplexing are equivalent to the possible $\psi(\boldsymbol{r})$ for the convex relaxation presented in Secs. \ref{sec:probeinter} and \ref{sec:convexrelax}.

Based on this mapping, the results of Sec. \ref{sec:convexrelax} imply that the above sensory multiplexing protocol achieves a fractional uncertainty that saturates to a constant value $\delta\lambda_0/\lambda_0 \approx \eta / \sqrt{\pi}$ in the asymptotic limit $k_{\mathrm{max}}$ of fine resolution. This saturation occurs because different boundary probes are correlated via their overlapping probe intensities $\psi_k(\boldsymbol{r})$ in the interior, which partially censors the information that the sensor can extract from the exterior. Thus, the sensor must account for these correlations in order to access the full extent of the information available from the exterior. To that end, the sensor can adjust each estimator to nullify their effective probe intensities in the interior. Specifically, the sensor can perform sensory multiplexing using the adjusted estimators $\hat{\zeta}_{0,k}$ given by the weighted sum of each estimator $\hat{\lambda}_{0,k}$ and an appropriately chosen companion estimator $\hat{\tilde{\lambda}}_{0,k}$:

\begin{equation} \label{eq:adjlam}
\hat{\zeta}_{0,k} = q_k \hat{\lambda}_{0,k} + \tilde{q}_k \hat{\tilde{\lambda}}_{0,k}.
\end{equation}

\noindent Imposing the additional constraint $q_k + \tilde{q}_k = 1$ ensures that the adjusted estimator $\hat{\zeta}_{0,k}$ is unbiased. Here, the companion estimator $\hat{\tilde{\lambda}}_{0,k}$ is given by:

\begin{equation}
\hat{\tilde{\lambda}}_{0,k} = \frac{\int \tilde{\psi}_k(\boldsymbol{r}) \lambda(\boldsymbol{r}) d\boldsymbol{r} }{ \int \tilde{\psi}_k(\boldsymbol{r}) d\boldsymbol{r}},
\end{equation}

\noindent where $\tilde{\psi}_k(\boldsymbol{r}) \equiv \nabla \tilde{V}_{f,k}(\boldsymbol{r}) \cdot \nabla \tilde{V}_{w,k}(\boldsymbol{r})$ is the probe intensity for the companion probe $k$. To cancel out the interior, we take $\tilde{V}_{f,k}(\boldsymbol{r})$ and $\tilde{V}_{w,k}(\boldsymbol{r})$ to be given by Eqs. (\ref{eq:companion1}) and (\ref{eq:companion2}) in the main text. These probe potentials can be generated by the probe fields $\tilde{f}_k(\boldsymbol{r}) = \nabla^2 \tilde{V}_{f,k}(\boldsymbol{r})$ and $\tilde{w}_k(\boldsymbol{r}) = \nabla^2 \tilde{V}_{w,k}(\boldsymbol{r})$, and they combine to yield the following probe intensities $\tilde{\psi}_k(\boldsymbol{r})$:

\begin{equation} \label{eq:proberho}
\tilde{\psi}_k(\boldsymbol{r})  = \begin{cases}
2k \left( \frac{r}{a}\right)^{2k-2}, & \text{$r<a$}.\\
0, & \text{$r>a$}.
\end{cases}
\end{equation}

\noindent Inserting the probe intensities in Eqs. (\ref{eq:psiint4}) and (\ref{eq:proberho}) into the estimators in Eq. (\ref{eq:adjlam}) results in:

\begin{equation}
\hat{\zeta}_{0,k} =  \frac{\int \Psi_k(\boldsymbol{r}) \lambda(\boldsymbol{r}) d\boldsymbol{r}}{\int \Psi_k(\boldsymbol{r}) d\boldsymbol{r}},
\end{equation}

\noindent where the effective probe intensities $\Psi_k(\boldsymbol{r})$ are given by:

\begin{equation} 
\Psi_k(\boldsymbol{r})  \sim \begin{cases}
q_k k\left( \frac{r}{a}\right)^{2k-2} + \tilde{q}_k 2k \left( \frac{r}{a}\right)^{2k-2} , & \text{$r<a$}.\\
q_k k\left( \frac{r}{a}\right)^{-2k-2}, & \text{$r>a$}.
\end{cases}
\end{equation}

\noindent Thus, to nullify the interiors of $\Psi_k(\boldsymbol{r})$, we must have $q_k + 2\tilde{q}_k = 0$. Combining this equation with the constraint for unbiasedness leads to $q_k = 2$ and $\tilde{q}_k = -1$ for all values of $k$. Inserting these values into the above equation results in the paired probes referred to in Sec. \ref{sec:greatjob} of the main text:

\begin{equation} \label{eq:pairedprobepsi}
\Psi_k(\boldsymbol{r})  \sim \begin{cases}
0, & \text{$r<a$}.\\
2 k \left( \frac{r}{a}\right)^{-2k-2}, & \text{$r>a$}.
\end{cases}
\end{equation}

\noindent Finally, the sensor may extract additional information by uniformly sampling the material constant field in its interior. To do so, the sensor may adjust the coefficient $\tilde{q}_1$. In this case, the optimal all-inclusive effective probe intensity $\Psi(\boldsymbol{r})$ is given by:

\begin{equation} \label{eq:allinclusive1}
\Psi(\boldsymbol{r}) = \sum_{k=1}^{k_{\mathrm{max}}} p_k \frac{\psi_k(\boldsymbol{r})}{\int \psi_k(\boldsymbol{r}) d\boldsymbol{r}} + \sum_{k=1}^{k_{\mathrm{max}}} \tilde{p}_k \frac{\tilde{\psi}_k(\boldsymbol{r})}{\int \tilde{\psi}_k(\boldsymbol{r}) d\boldsymbol{r}},
\end{equation}

\noindent for appropriate values of $p_k$ and $\tilde{p}_k$. Alternatively, the sensor may incorporate an unpaired probe $\psi_0(\boldsymbol{r})$ given by Eq. (\ref{eq:psiint}) into the sensory multiplexing protocol. This unpaired probe is associated with the following optimal estimator $\hat{\zeta}_{0,0} $:

\begin{equation}
\hat{\zeta}_{0,0} = \frac{\int \psi_0(\boldsymbol{r}) \lambda(\boldsymbol{r}) d\boldsymbol{r}}{\int \psi_0(\boldsymbol{r}) d\boldsymbol{r}}.
\end{equation}

\noindent Combining this unpaired probe with the paired probes described by Eq. (\ref{eq:pairedprobepsi}) results in the following optimal all-inclusive effective probe intensity $\Psi(\boldsymbol{r})$:

\begin{equation} \label{eq:allinclusive2}
\Psi(\boldsymbol{r}) = p_0 \frac{ \psi_0(\boldsymbol{r})}{\int \psi_0(\boldsymbol{r}) d\boldsymbol{r}} + \sum_{k=1}^{k_{\mathrm{max}}} p_k \frac{\Psi_k(\boldsymbol{r})}{\int \Psi_k(\boldsymbol{r}) d\boldsymbol{r}},
\end{equation}

\noindent where $p_k = \sum_{l} C^{-1}_{kl}$ with $C_{kl} \equiv \langle (\hat{\zeta}_{0,k} - \lambda_0) (\hat{\zeta}_{0,l} - \lambda_0) \rangle$. The all-inclusive effective probe intensities given by Eqs. (\ref{eq:allinclusive1}) and (\ref{eq:allinclusive2}) are equivalent in that their configurations are identical and that they result in the same fractional uncertainty. However, their sensory multiplexing protocols are physically distinct in that the latter contains one additional probe. In Sec. \ref{sec:smproof}, we demonstrate that this all-inclusive effective probe intensity exhaustively probes the full extent of the information available to the sensor.

\section{A fundamental, physical limit to the fractional uncertainty for the two-dimensional elastic sheet} \label{sec:smproof}

In this section, we prove that the adjusted sensory multiplexing protocol presented in Sec. \ref{sec:greatjob} of the main text achieves the smallest possible fractional uncertainty $\delta \lambda_0 / \lambda_0$ that is obtained among all sensory multiplexing protocols that employ probe fields confined to a region $r \le a$. To do so, we consider a sensory multiplexing protocol containing an arbitrary number of arbitrarily complicated probes. To determine the smallest $\delta \lambda_0 / \lambda_0$ for such a protocol, we start by imagining a convex relaxation of $\Psi(\boldsymbol{r})$ that generalizes the convex relaxation presented in Sec. (\ref{sec:probeinter}) for a single boundary probe. Specifically, we expand the space of possible $\Psi(\boldsymbol{r})$ to allow an arbitrary configuration in the interior combined with any configuration in the exterior that can be generated by the convex relaxation presented in Sec. (\ref{sec:probeinter}). It follows that the minimum fractional uncertainty $\delta \lambda_{0,\mathrm{low}} / \lambda_0$ for this convex relaxation provides a theoretical lower bound on $\delta\lambda_0 / \lambda_0$. Thus, provided that we can construct physical configurations of the probe potentials that realize this lower bound, $\delta \lambda_{0,\mathrm{low}} / \lambda_0$ provides a fundamental, physical limit to the fractional uncertainty $\delta \lambda_0 / \lambda_0$.

We determine $\delta \lambda_{0,\mathrm{low}} / \lambda_0$ by first separating the action $S$ for the variance $\delta\lambda_0^2$ into the sum $S = S_{\mathrm{int}} + S_{\mathrm{ext}}$ of contributions from the interior $\mathcal{R}_{\mathrm{int}}$ and the exterior $\mathcal{R}_{\mathrm{ext}}$. The contribution $S_{\mathrm{int}}$ is minimized by a uniform $\Psi(\boldsymbol{r}) = \gamma$ in $\mathcal{R}_{\mathrm{int}}$, as we found in Sec. \ref{sec:probewinkler}. For the exterior, the optimal configuration of $\Psi(\boldsymbol{r})$ is obtained by minimizing $S_{\mathrm{ext}}$ over the coefficients of the boundary modes. To that end, our treatment of the convex relaxation in Sec. \ref{sec:convexrelax} implies that the optimal $\Psi(\boldsymbol{r})$ in the exterior must only receive contributions from the diagonal terms of $\mathcal{M}^{(-)}_{jk}$. These contributions can be mapped onto the paired probes $\Psi_k(\boldsymbol{r})$ in Sec. \ref{sec:smsheet} via Eq. (\ref{eq:pkpkkmap}). Finally, we observe that the all-inclusive effective probe intensity $\Psi(\boldsymbol{r})$ described in Sec. \ref{sec:greatjob} of the main text simultaneously minimizes $S_{\mathrm{int}}$ and $S_{\mathrm{ext}}$ subject to the constraint $\int \Psi(\boldsymbol{r}) d\boldsymbol{r} = 1$. Thus, the all-inclusive effective probe intensity $\Psi(\boldsymbol{r})$ provides an estimate of $\lambda_0$ with the smallest physically possible fractional uncertainty $\delta \lambda_0 / \lambda_0$. This proof generalizes to $D=3$ in a straightforward manner.

\section{Sensory multiplexing for a three-dimensional elastic medium} \label{sec:smsolid}

In this section, we quantify the precision of a sensor that can perform multiple probes of a three-dimensional, elastic medium. For simplicity, we constrain the medium to deform as a scalar $u(\boldsymbol{r})$ at each point in space, analogous to our treatment of the two-dimensional elastic sheet. Physically, this medium corresponds to an anisotropic elastic solid constrained to deform along a single direction. The internal energy of such an elastic solid is given by:

\begin{equation} \label{eq:esheeta}
E = \frac{1}{2} \int \lambda(\boldsymbol{r}) \nabla u(\boldsymbol{r}) \cdot \nabla  u(\boldsymbol{r}) d\boldsymbol{r}.
\end{equation}

\noindent Here, as for the Winkler foundation and the elastic sheet, we take $\lambda(\boldsymbol{r})$ to be a Gaussian random field with mean $\lambda_0$, variance $\Delta_{\lambda} \ll \lambda_0^2$, and spatial correlations over a scale $\xi$. As before, we take the sensor to interact with the medium within a radius $a$ by first applying a stimulus field $f(\boldsymbol{r})$ as in Eq. (\ref{eq:fspring}), and then measuring an integrated response $m$ as in Eq. (\ref{eq:mspring}).

In what follows, we will determine the fractional uncertainty for a sensory multiplexing protocol applied to this elastic medium in the asymptotic limit of fine spatial resolution. Motivated by the results in the main text for the case of the elastic sheet, we take each probe $i$ to apply the following probe fields:

\begin{equation}
f_i(\boldsymbol{r}) \sim \delta (r - a) Y_{\ell_i m_i}(\theta, \varphi),
\end{equation}

\begin{equation}
w_i(\boldsymbol{r}) \sim \delta (r - a) Y^*_{\ell_i m_i}(\theta, \varphi),
\end{equation}

\noindent where $Y_{\ell_i, m_i}$ are spherical harmonics of degree $\ell_i$ and order $m_i$ and $Y^*_{\ell_i, m_i}$ are their complex conjugates. To be concrete, we choose the prefactors of the probe fields such that the probe potentials are given by:

\begin{equation}
V_{f,i}^{\mathrm{(+)}}(\boldsymbol{r}) = \frac{r^{\ell}Y_{\ell_i m_i}(\theta, \varphi)}{(2\ell+1)a^{\ell-1}}.
\end{equation}

\begin{equation}
V_{w,i}^{\mathrm{(+)}}(\boldsymbol{r}) = \frac{r^{\ell}Y^*_{\ell_i m_i}(\theta, \varphi)}{(2\ell+1)a^{\ell-1}}.
\end{equation}

\begin{equation}
V_{f,i}^{\mathrm{(-)}}(\boldsymbol{r}) = \frac{a^{\ell+2}Y_{\ell_i m_i}(\theta, \varphi)}{(2\ell+1)r^{\ell+1}}.
\end{equation}

\begin{equation}
V_{w,i}^{\mathrm{(-)}}(\boldsymbol{r}) = \frac{a^{\ell+2}Y^*_{\ell_i m_i}(\theta, \varphi)}{(2\ell+1)r^{\ell+1}},
\end{equation}

\noindent in the interior and exterior, respectively. Moreover, we assume that the sensor executes such probes for all possible values of $m$ and $\ell$ up to a maximum degree $\ell_{\mathrm{max}}$. For this sensory geometry, we define the sensor resolution $d$ to be inversely proportional to $\ell_{\mathrm{max}}$:

\begin{equation}
d \sim \frac{a}{\ell_{\mathrm{max}}}.
\end{equation}

\noindent As before for the two-dimensional elastic sheet, the MVUE of $\lambda_0$ for the above sensory multiplexing protocol is given by the best linear unbiased sum of the estimators $\hat{\lambda}_{0,i}$ of individual probes:

\begin{equation} 
\hat{\lambda}_0 = \sum_i p_i \hat{\lambda}_{0,i},
\end{equation}

\noindent where the estimator weights $p_i$ are the following normalized sums over the rows of the inverse $C_{ij}^{-1}$ of the covariance matrix:

\begin{equation} 
p_i = \frac{\sum_{j}C_{ij}^{-1}}{ \sum_{ij} C_{ij}^{-1} }.
\end{equation}

\noindent The covariance matrix $C_{ij}$ is defined by Eq. (\ref{eq:cmn}). The variance $\delta \lambda_0^2$ of the estimator $\hat{\lambda}_{0}$ is given by:

\begin{equation} \label{eq:cpipi}
\delta \lambda_0^2 = \sum_{ij} p_i p_j C_{ij}.
\end{equation}

\noindent To calculate the covariance matrix, we insert the probe potentials into Eq. (\ref{eq:cmn}), which results in:

\begin{equation} \label{eq:cpcovfarout}
C_{ij} = \Delta_{\lambda} \xi^3 \left( \int_{\mathcal{R}_{\mathrm{int}}} \frac{\psi_i^{\mathrm{(+)}}(\boldsymbol{r})  \psi_j^{\mathrm{(+)}}(\boldsymbol{r})      }{s_i s_j} d\boldsymbol{r}  +  \int_{\mathcal{R}_{\mathrm{ext}}} \frac{\psi_i^{\mathrm{(-)}}(\boldsymbol{r})  \psi_j^{\mathrm{(-)}}(\boldsymbol{r})      }{s_i s_j} d\boldsymbol{r}    \right),
\end{equation}

\noindent where

\begin{equation}
\psi_i^{\mathrm{(\pm)}}(\boldsymbol{r}) = \nabla V_{w,i}^{\mathrm{(\pm)}}(\boldsymbol{r}) \cdot \nabla V_{f,i}^{\mathrm{\pm()}}(\boldsymbol{r}),
\end{equation}

\noindent and

\begin{equation}
s_i = \int_{\mathcal{R}_{\mathrm{int}}} \psi_i^{(+)}(\boldsymbol{r}) d\boldsymbol{r} + \int_{\mathcal{R}_{\mathrm{ext}}} \psi_i^{(-)}(\boldsymbol{r}) d\boldsymbol{r},
\end{equation}

\noindent are the normalizing constants given by:

\begin{equation}
s_i  = \frac{a^3}{2\ell+1}.
\end{equation}

\noindent Inserting the covariance matrix given by Eq. (\ref{eq:cpcovfarout}) into Eq. (\ref{eq:cpipi}) results in:

\begin{equation}
\delta \lambda_0^2 = \Delta_{\lambda} \xi^3  \sum_{ij} p_i p_j  \left( \int_{\mathcal{R}_{\mathrm{(+)}}} \frac{\psi_i^{\mathrm{(+)}}(\boldsymbol{r})  \psi_j^{\mathrm{(+)}}(\boldsymbol{r})      }{s_i s_j} d\boldsymbol{r}  +  \int_{\mathcal{R}_{\mathrm{ext}}} \frac{\psi_i^{\mathrm{(-)}}(\boldsymbol{r})  \psi_j^{\mathrm{(-)}}(\boldsymbol{r})      }{s_i s_j} d\boldsymbol{r}    \right),
\end{equation}

\noindent which can be separated into the sum $\delta\lambda_0^2 = \delta\lambda_{0,\mathrm{int}}^2 + \delta\lambda_{0,\mathrm{ext}}^2 $ of contributions from the interior and the exterior:

\begin{equation}
\delta\lambda_{0,\mathrm{int}}^2 = \Delta_{\lambda} \xi^3  \sum_{ij} p_i p_j  \int_{\mathcal{R}_{\mathrm{int}}} \frac{\psi_i^{\mathrm{(+)}}(\boldsymbol{r})  \psi_j^{\mathrm{(+)}}(\boldsymbol{r})      }{s_i s_j} d\boldsymbol{r} ,
\end{equation}

\begin{equation}
\delta\lambda_{0,\mathrm{ext}}^2 = \Delta_{\lambda} \xi^3  \sum_{ij} p_i p_j  \int_{\mathcal{R}_{\mathrm{ext}}} \frac{\psi_i^{\mathrm{(-)}}(\boldsymbol{r})  \psi_j^{\mathrm{(-)}}(\boldsymbol{r})      }{s_i s_j} d\boldsymbol{r}.
\end{equation}

\noindent To evaluate these integrals, we first consider the contribution $\delta\lambda_{0,\mathrm{ext}}^2 $ from the exterior. The above equation can be expressed in terms of the probe potentials as follows:

\begin{equation}
\delta\lambda_{0,\mathrm{ext}}^2 = \Delta_{\lambda} \xi^3 \sum_{\ell_i=0 }^{\ell_{\mathrm{max}}} \sum_{\ell_j=0 }^{\ell_{\mathrm{max}}} \sum_{m_i=-\ell_i }^{\ell_i} \sum_{m_j=-\ell_j }^{\ell_j} p_{i} p_{j} \int_{\mathcal{R}_{\mathrm{ext}}}  \frac{  a^{ 2\ell_i + 1}}{ (2 \ell_i + 1)}   \nabla \left( \frac{Y_{\ell_i m_i}}{r^{\ell_i + 1}}\right) \cdot \nabla \left( \frac{Y^*_{\ell_i m_i}}{r^{\ell_i + 1}}\right)  \frac{ a^{ 2\ell_j + 1}}{ (2 \ell_j + 1)}   \nabla \left( \frac{Y_{\ell_j m_j}}{r^{\ell_j + 1}}\right) \cdot \nabla \left( \frac{Y^*_{\ell_j m_j}}{r^{\ell_j + 1}}\right) d \boldsymbol{r}.
\end{equation}

\noindent We swap the order of the sums and the integration to obtain:

\begin{equation}\label{eq:varfac}
\delta \lambda_0^2 = \Delta_{\lambda} \xi^3 \int_{\mathcal{R}_{\mathrm{ext}}} \left( \sum_{\ell_i=0 }^{\ell_{\mathrm{max}}}  \sum_{m_i=-\ell_i }^{\ell_i}    \frac{ p_{i} a^{ 2\ell_i + 1}}{ (2 \ell_i + 1)}   \nabla \left( \frac{Y_{\ell_i m_i}}{r^{\ell_i + 1}}\right) \cdot \nabla \left( \frac{Y^*_{\ell_i m_i}}{r^{\ell_i + 1}}\right) \right)^2 d \boldsymbol{r}.
\end{equation}

\noindent To compute the sum over the spherical harmonic orders $m_i$, we must know the values of $p_i$. Numerical minimization of the variance indicates that the values of these coefficients are independent of the spherical harmonic orders $m_i$. Using this ansatz, we can express the values of these coefficients as:

\begin{equation}
p_i = \frac{p_{\ell_i}}{2\ell_i + 1},
\end{equation}

\noindent where $p_{\ell_i}$ is a constant that depends on the degree $\ell_i$ of the probe $i$ and $\sum_i p_{\ell_i} = 1$. Inserting this expression for the weights in Eq. (\ref{eq:varfac}) yields:

\begin{equation}\label{eq:varfac2}
\delta\lambda_{0,\mathrm{ext}}^2 = \Delta_{\lambda} \xi^3 \int_{\mathcal{R}_{\mathrm{ext}}} \left( \sum_{\ell_i=0 }^{\ell_{\mathrm{max}}}  \sum_{m_i=-\ell_i }^{\ell_i}    \frac{ p_{\ell_i} a^{ 2\ell_i + 1}}{ (2 \ell_i + 1)^2}   \nabla \left( \frac{Y_{\ell_i m_i}}{r^{\ell_i + 1}}\right) \cdot \nabla \left( \frac{Y^*_{\ell_i m_i}}{r^{\ell_i + 1}}\right) \right)^2 d \boldsymbol{r}.
\end{equation}

\noindent To proceed, we employ the closure relationship for the sum of the spherical harmonic orders \cite{Jackson1998}, which implies the following identity:

\begin{equation}
\sum_{m}  \nabla \left( f(r)Y_{\ell m}   \right) \cdot  \nabla \left(  f(r)Y^*_{\ell m} \right) = \frac{2\ell +1}{4\pi} \left( \frac{\ell (\ell+1) f(r)^2}{r^2} + f'(r)^2 \right).
\end{equation}

\noindent Using this identity, we take the sum over the orders $m_i$ in Eq. (\ref{eq:varfac2}) to find:

\begin{equation}
\delta\lambda_{0,\mathrm{ext}}^2 = \Delta_{\lambda} \xi^3 \int_{\mathcal{R}_{\mathrm{ext}}}  \left( \sum_{\ell_i } p_{\ell_i} \frac{(\ell_i+1 )a^{2\ell_i+1}}{4\pi r^{2\ell_i+4}} \right)^2  d\boldsymbol{r}.
\end{equation}

\noindent We now expand the sum and perform the integral to obtain:

\begin{equation}
\delta\lambda_{0,\mathrm{ext}}^2 =  \frac{\Delta_{\lambda}}{4\pi} \left(\frac{\xi}{a}\right)^3    \sum_{\ell_i \ell_j}  p_{\ell_i} p_{\ell_j} \frac{ (\ell_i + 1) (\ell_j + 1)}{2 \ell_i + 2 \ell_j + 5} .
\end{equation}

\noindent A similar calculation can be performed for the interior, which yields:

\begin{equation}
\delta\lambda_{0,\mathrm{int}}^2 =  \frac{\Delta_{\lambda}}{4\pi} \left(\frac{\xi}{a}\right)^3    \sum_{\ell_i \ell_j}  p_{\ell_i} p_{\ell_j} \frac{  \ell_i \ell_j}{2 \ell_i + 2 \ell_j -1} .
\end{equation}

\noindent Adding up the contributions $\delta\lambda_{0,\mathrm{int}}^2 $ and $\delta\lambda_{0,\mathrm{ext}}^2$ from the interior and the exterior results in:

\begin{equation}
\delta\lambda_{0}^2 =  \frac{\Delta_{\lambda}}{4\pi} \left(\frac{\xi}{a}\right)^3    \sum_{\ell_i \ell_j}  p_{\ell_i} p_{\ell_j} \tilde{C}_{ij} ,
\end{equation}

\noindent where $\tilde{C}_{ij}$ is a dimensionless matrix given by:

\begin{equation}
\tilde{C}_{ij} =  \frac{  \ell_i \ell_j}{2 \ell_i + 2 \ell_j -1} + \frac{ (\ell_i + 1) (\ell_j + 1)}{2 \ell_i + 2 \ell_j + 5}.
\end{equation}

\noindent In terms of this matrix, the variance $\delta \lambda_0^2$ is given by:

\begin{equation}
\delta \lambda_0^2 =  \frac{\Delta_{\lambda}}{4\pi} \left(\frac{\xi}{a}\right)^3  \left( \sum_{ij} \tilde{C}_{ij}^{-1} \right)^{-1} .
\end{equation}

\noindent As before for the two-dimensional elastic sheet, we find that the contributions from the interior introduce correlations among the probes that limit the amount of information that can be extracted. To remove these unnecessary correlations, we consider the companion probes given by:

\begin{equation}
V_{f,\ell}(\boldsymbol{r}) \sim a-r,
\end{equation}

\begin{equation}
V_{w,\ell}(\boldsymbol{r}) \sim \left(\frac{r}{a}\right)^{2\ell-2} - 1.
\end{equation}

\noindent We combine these probes with those of the original protocol, as in Sec. \ref{sec:smsheet}. This yields the following adjusted matrix $\tilde{\tilde{C}}_{ij}$:

\begin{equation} \label{eq:ctt}
\tilde{\tilde{C}}_{ij} = \frac{ (2\ell_i + 1) (2\ell_j + 1)}{2 \ell_i + 2 \ell_j + 5},
\end{equation}

\noindent and a variance given by:

\begin{equation}
\delta \lambda_0^2 = \frac{\Delta_{\lambda}}{4\pi} \left(\frac{\xi}{a}\right)^3 \left( \sum_{ij} \tilde{\tilde{C}}_{ij}^{-1} \right)^{-1} .
\end{equation}

\noindent Inserting the adjusted matrix $\tilde{\tilde{C}}_{ij}$ given by Eq. (\ref{eq:ctt}) into the above equation results in:

\begin{equation} \label{eq:lamyup}
\delta \lambda_0^2 = \frac{\Delta_{\lambda}}{4\pi} \left(\frac{\xi}{a}\right)^3  \left( \sum_{k=1}^{\ell_{\mathrm{max}}+1} \Upsilon(k)   \right)^{-1},
\end{equation}

\noindent where $\Upsilon(k)$ is given by:

\begin{equation}
\Upsilon(k) = (4k+5) \frac{(k+1)!(k+1)!}{(k+1/2)!(k+1/2)!}.
\end{equation}

\noindent In the limit $k \rightarrow \infty$, this function scales as $\Upsilon(k) \sim k^2$ (by Stirling's approximation). Thus, for a large maximum degree $\ell_{\mathrm{max}} \rightarrow \infty$, the sum in Eq. (\ref{eq:lamyup}) approaches:

\begin{equation}
\delta \lambda_0^2 \sim \frac{\Delta_{\lambda}}{4\pi} \left(\frac{\xi}{a}\right)^3  \left(  \sum_{k=\ell_{\mathrm{0}}}^{\ell_{\mathrm{max}}+1} k^2  \right)^{-1},
\end{equation}

\noindent where $\ell_0 \gg 1$. We compute the sum of these consecutive squares to find:

\begin{equation}
\delta \lambda_0^2 \sim \frac{\Delta_{\lambda}}{4\pi} \left(\frac{\xi}{a}\right)^3  \left( \frac{(2\ell_{\mathrm{max}}+1)\ell_{\mathrm{max}}(\ell_{\mathrm{max}}+1)}{6} - \frac{(2\ell_0+1)\ell_0(\ell_0+1)}{6}  \right)^{-1}.
\end{equation}

\noindent Finally, we take the limit $\ell_{\mathrm{max}} \gg \ell_0$ to obtain the following scaling for the fractional uncertainty:

\begin{equation}
\frac{\delta \lambda_0}{\lambda_0 } \sim \left( \frac{\Delta_{\lambda} }{ \lambda_0^2} \right)^{1/2} \left( \frac{ d }{ a } \right)^{3/2} \left( \frac{ \xi }{ a } \right)^{3/2}    ,
\end{equation}

\noindent which matches Eq. (\ref{eq:funcert}) in the main text for $D=3$.

\section{Sensory multiplexing is robust to the omission of modes} \label{sec:smomit}

In the main text, we considered sensory multiplexing protocols that harnessed all possible mode pairs up to a maximum mode number $k_{\mathrm{max}}$. How does the precision of the sensor change if this assumption is violated? To gain insight into this question, we consider a sensor that executes the paired probes in Sec. (\ref{sec:greatjob}) starting from an initial mode number $k_{\mathrm{min}}$ up to a maximum mode number $k_{\mathrm{max}}$. For this sensory multiplexing protocol, the covariance matrix is again given by Eq. (\ref{eq:cmnsol}). Inserting the inverse of this matrix into Eq. (\ref{eq:cvar}) and taking the sums to range over the included probes results in the variance:

\begin{equation}
\delta \lambda_0^2 = 2\Delta_{\lambda} \left( \frac{\xi}{a} \right)^2 \frac{k_{\mathrm{min}}^2}{(k_{\mathrm{max}} + k_{\mathrm{min}} + 1)(k_{\mathrm{max}} - k_{\mathrm{min}} + 1)}.
\end{equation}

\noindent This variance increases with $k_{\mathrm{min}}$, which indicates that the precision of the sensor worsens as mode pairs are omitted. Nevertheless, in the limit $k_{\mathrm{max}} \gg k_{\mathrm{min}}$, the scaling of the fractional uncertainty with the sensor's resolution is again given by Eq. (\ref{eq:funcert}). Moreover, we have explored variants of the above protocol that consist of omitting intermediate mode pairs, and found that they also obey the scaling in Eq. (\ref{eq:funcert}) for $k_{\mathrm{max}} \gg k_{\mathrm{min}}$. Taken together, our results suggest that the details of the measurement protocol do not affect the scaling of $\delta\lambda_0 / \lambda_0$ with $d$, provided that the sensor probes a sufficiently large number of mode pairs.

\section{A numerical lower bound on $\delta\lambda_0 / \lambda_0$ for volume probes} \label{sec:bulknum}

In the main text, we showed how interferences between the modes contained in the probe fields can fundamentally limit the precision of a sensor. In particular, we found that a sensor limited to applying a single probe on its boundary can never attain the smallest possible fractional uncertainty for a sensor that can perform multiple probes. To what extent do these interference effects limit the precision of a sensor that can apply a single, arbitrary probe within its volume? To gain insight into this question, we extended our numerical approach to account for such volume probes.

Probe fields containing bulk modes can potentially provide a number $\sim k_{\mathrm{max}}^{16}$ of contributions to $\delta \lambda^2_{0}$. This rapid scaling drastically limits the scope of conventional numerical minimization. Therefore, to maximize the reach of our computational capabilities, we considered a constraint relaxation of $\delta \lambda_{0} / \lambda_0$ that allows us to treat the interior and exterior of the sensor separately. Specifically, we separate the action $S$ for the variance $\delta\lambda_0^2$ into the sum $S = S_{\mathrm{int}} + S_{\mathrm{ext}}$ of contributions from the interior $\mathcal{R}_{\mathrm{int}}$ and the exterior $\mathcal{R}_{\mathrm{ext}}$. We then minimize $S_{\mathrm{int}}$ and $S_{\mathrm{ext}}$ individually, disregarding the constraint that the probe potentials must be continuous across the sensor's boundary $\mathcal{B}$. It follows that the sum of the minima of $S_{\mathrm{int}}$ and $S_{\mathrm{ext}}$ provide a lower bound $\delta \lambda^2_{0,\mathrm{low}}$ on the minimum of $S$ (see Sec. \ref{sec:boundvar}).

The contribution $S_{\mathrm{int}}$ is minimized by a uniform probe intensity $\psi(\boldsymbol{r}) = \gamma$ in $\mathcal{R}_{\mathrm{int}}$, as we found in Sec. \ref{sec:probewinkler} of the main text. For the exterior, the probe potentials are completely determined by their values on the boundary, which allowed us to employ the same numerical optimization scheme as for the boundary probes (see Sec. \ref{sec:snum}). For all values of $k_{\mathrm{max}} > 1$, we found that the resulting probe potentials were dominated by the dipole-dipole and quadrupole-quadrupole pairs, with higher order modes observed for $k_{\mathrm{max}} > 8$ (see Fig. 3 of the main text). Moreover, in this case, the theoretical lower bound $\delta \lambda_{0,\mathrm{low}} / \lambda_0$ on the fractional uncertainty does not provide a close match to the fractional uncertainties obtained for sensory multiplexing. This discrepancy suggests that even if a sensor is capable of applying an arbitrary pair of probe fields within its volume, interferences between modes significantly restrict the amount of information that the sensor can glean in comparison to sensory multiplexing.

%\setcitestyle{super}

\section{The probe intensity for continuum elasticity} \label{sec:pice}

In this section, we determine the probe intensity for the elastic medium in our model of cellular mechanosensing. Taking the variation of the internal energy given by Eq. (\ref{eq:intelsol}) with respect to the deformation field $u_i$ results in the following constitutive relation:

\begin{equation}
\delta_{i,k} \partial_j (\mu \partial_j u_i) + \partial_i (\mu \partial_k u_i) + c_0 \partial_k (\mu \partial_i u_i)  = f_k.
\end{equation}

\noindent To determine the probe intensity, we expand the deformation field to leading order in $\delta \lambda(\boldsymbol{r})$. This approach yields an approximate deformation field given by the sum of a zeroth order deformation field $u_i^{(0)}$ and a first order deformation field $u_i^{(1)}$. We solve for the zeroth order integrated measurement $m^{(0)}$ by inverting the above constitutive relation to find:

\begin{equation}
u^{(0)}_i = \frac{1}{\mu_0} \int G_{ik} f_k d\boldsymbol{r}
\end{equation}

\noindent where $G_{ik}$ is the response function defined by:

\begin{equation}
(\delta_{i,k} \partial_j \partial_j + \partial_i \partial_k + c_0 \partial_i \partial_k ) G_{ia} = \delta_{k,a}.
\end{equation}

\noindent The leading order integrated measurement $m^{(0)}$ is given by:

\begin{equation}
m^{(0)} = \int w_i u^{(0)}_i d\boldsymbol{r}.
\end{equation}

\begin{equation}
m^{(0)} = \frac{1}{\mu_0} \int w_i V_{f,i} d\boldsymbol{r},
\end{equation}

\noindent where $V_{f,i}$ is the stimulus potential:

\begin{equation}
V_{f,i} = \int G_{ik} f_k d\boldsymbol{r}.
\end{equation}

\noindent Equivalently, the stimulus potential is also defined by the following equation:

\begin{equation}
(\delta_{i,k} \partial_j \partial_j +  \partial_i \partial_k + c_0  \partial_k  \partial_i ) V_{f,i} = f_k.
\end{equation}

\noindent Similarly, we define a weight potential:

\begin{equation}
V_{w,i} = \int G_{ik} w_kd\boldsymbol{r}.
\end{equation}

\begin{equation}
(\delta_{i,k} \partial_j \partial_j + \partial_i \partial_k + c_0  \partial_k \partial_i) V_{w,i} = w_k.
\end{equation}

\noindent Inserting the expression for the weight potential into the leading order integrated measurement $m^{(0)}$ yields:

\begin{equation}
m^{(0)} = \frac{1}{\mu_0} \int (\delta_{i,k} \partial_j \partial_j + \partial_k \partial_i + c_0  \partial_i \partial_k ) V_{w,k} V_{f,i} d\boldsymbol{r}.
\end{equation}

\noindent Upon integrating this expression by parts, we find:

\begin{equation} \label{eq:zeroorder}
m^{(0)} = \frac{1}{\mu_0} \int (\partial_j V_{f,i} \partial_j V_{w,i} +  \partial_i V_{f,k} \partial_k V_{w,i} + c_0 \partial_i V_{f,i} \partial_k V_{w,k}) d\boldsymbol{r}.
\end{equation}

\noindent We now turn to the first order integrated measurement $m^{(1)}$. To leading order in $\delta \mu$, the first-order deformation field $u_i^{(1)}$ is:

\begin{equation}
u_i^{(1)} = \frac{-1}{\mu_0^2} \int G_{ik} \left( \delta_{i,k} \partial_j (\delta \mu \partial_j V_{f,i}) + \partial_i (\delta \mu \partial_k V_{f,i} ) + c_0 \partial_k (\delta \mu \partial_i V_{f,i} ) \right) d\boldsymbol{r}.
\end{equation}

\noindent Thus, the first-order integrated measurement $m^{(1)}$ is:

\begin{equation}
m^{(1)} = \int w_i u_i^{(1)} d\boldsymbol{r} = \frac{-1}{\mu_0^2} \int \left(  V_{w,i} \partial_j (\delta \mu \partial_j V_{f,i} )  +  V_{w,k} \partial_i (\delta \mu \partial_k V_{f,i} )  + c_0  V_{w,k} \partial_k (\delta \mu \partial_i V_{f,i} )   \right) d\boldsymbol{r}.
\end{equation}

\noindent We integrate by parts to find:

\begin{equation} \label{eq:firstorder}
m^{(1)} = \frac{-1}{\mu_0^2} \int \delta \mu \left(  \partial_j V_{w,i}   \partial_j V_{f,i}   +  \partial_i V_{w,k}   \partial_k V_{f,i}   + c_0 \partial_k   V_{w,k}   \partial_i V_{f,i}   \right) d\boldsymbol{r}.
\end{equation}

\noindent Adding together Eqs. (\ref{eq:zeroorder}) and (\ref{eq:firstorder}) results in the following integrated measurement $m \equiv m^{(0)} + m^{(1)}$:

\begin{equation}
m = \int \left( \frac{1}{\mu_0} - \frac{ \delta \mu(\boldsymbol{r})} {\mu_0^2} \right) \psi(\boldsymbol{r}) d\boldsymbol{r},
\end{equation}

\noindent where we have defined the probe intensity $\psi(\boldsymbol{r})$:

\begin{equation}
\psi(\boldsymbol{r}) = \partial_j V_{f,i} \partial_j V_{w,i} +  \partial_i V_{f,k} \partial_k V_{w,i} + c_0 \partial_i V_{f,i} \partial_k V_{w,k}.
\end{equation}

\noindent By analogy to the Winkler foundation, it follows that the sensor can obtain an unbiased estimate of ${\lambda}_0$ for the elastic sheet by inserting this probe intensity into Eq. (\ref{eq:haha}).

\section{Estimating the material parameters for a biopolymer network} \label{sec:params}

%\setcitestyle{numbers}

For our study of cellular mechanosensing, we determined the parameters of the elastic medium for a reconstituted collagen network, an \emph{in vitro} system that closely resembles \emph{in vivo} cellular environments \cite{Doyle2016,Zaman2006,Guo2013,Beroz2017}. Although many previous studies have measured the bulk mechanics of reconstituted collagen networks \cite{Janmey1983,Roeder2002,Knapp1997,Arevalo2010}, no existing studies have reported the parameters $\Delta_{\mu}$ and $\xi$ used in our model to characterize the local heterogeneity. Therefore, to determine both the bulk and local parameters, we fit our continuum model to the results of Ref. \cite{Beroz2017}, which reported a detailed characterization of the local response distribution for a reconstituted collagen network.

In this previous work, the authors inferred the local mechanical response of an experimental collagen network from a computational analysis of its structure. The collagen network was polymerized from a $c\sim \SI{0.2}{\microgram/\milli\liter}$ solution of collagen type-I monomers, and its resulting architecture was imaged using confocal microscopy. This architecture was then used as input to a discrete fiber network model. The parameters of this network model were determined by fitting to bulk rheology performed on the experimental network. Finally, the authors quantified the local mechanical response by simulating the response of the discrete network to localized force dipoles. From these simulations, the authors calculated the distribution of local stiffnesses, defined as the linear deformation response to a pair of equal-and-opposite forces acting on two vertices of the network separated by a given distance. Based on this analysis, we determined the parameters of our elastic medium as follows:

\begin{itemize}
	
	\item{The shear modulus $\mu_0$ was taken to be the value $\mu_0\simeq \SI{0.3}{\pascal}$ measured for the experimental network using bulk rheology.}
	
	\item{The Poisson's ratio $\sigma$ was not reported in Ref. \cite{Beroz2017}. Thus, we take its value to be $\sigma \simeq 0.4$, consistent with previous studies of the bulk response of collagen networks \cite{Knapp1997,Castro2016}.}
	
	\item{To determine the local variability of the material constant field $\Delta_{\mu}$, we compared the local stiffness distribution of the collagen network to the local response distribution for a measurement protocol in our continuum model that consists of a completely anisotropic force dipole. The deformation field produced by such a dipole is proportional to:
		\begin{equation} \label{eq:elasticdisp}
		V_{i}(\boldsymbol{r}) = G_{ij,k}(\boldsymbol{r}) P_{jk},
		\end{equation} 
		where $G_{ij,k}(\boldsymbol{r})$ is the gradient of the response function $G_{ij}(\boldsymbol{r})$ for a continuous elastic medium \cite{Landau}:
		\begin{equation}
		G_{ij}(\boldsymbol{r}) = \left[(3-4\sigma)\delta_{i,j} + \hat{r}_i \hat{r}_j \right] \frac{1}{r}.
		\end{equation}
		In this expression, $\hat{\boldsymbol{r}}$ is a unit vector oriented along $\boldsymbol{r}$, and $P_{jk}$ is the dipole moment tensor \cite{Clouet2018}. For a completely anisotropic dipole, this tensor can be expressed as:
		\begin{equation}
		P_{jk} = \delta_{1,j}\delta_{1,k}.
		\end{equation}
		In contrast to the force dipoles acting on the discrete network, a force dipole in the continuum limit induces diverging deformations at the points where the forces are applied. To account for these unphysical divergences, we take the measurement protocol to include a spherical cutoff region of radius $a$ equal to the length of the dipoles applied to the discrete network. Applying this cutoff to the deformation field in Eq. (\ref{eq:elasticdisp}) results in the following measurement protocols:
		\begin{equation} \label{eq:farpot}
		V_{f,i}(\boldsymbol{r}) \sim V_{w,i}(\boldsymbol{r}) \sim \begin{cases}
		\partial_k \left( \left[(3-4\sigma)\delta_{i,j} + \hat{r}_i \hat{r}_j \right] \frac{r^2}{a^3} \right)P_{jk}, & \text{$r<a$},\\
		\partial_k \left( \left[(3-4\sigma)\delta_{i,j} + \hat{r}_i \hat{r}_j \right] \frac{1}{r} \right)P_{jk}, & \text{$r>a$}.
		\end{cases}
		\end{equation}
		
		Using this measurement protocol, we determined $\Delta_{\mu}$ by computing the fractional uncertainty $\Delta_{\mu}/\mu_0$ in our continuum model via numerical integration and setting it equal to the corresponding fractional uncertainty found for the local response in Ref. \cite{Beroz2017}, i.e. the standard deviation of the local stiffness distribution divided by its mean. This comparison resulted in a value of $\Delta_{\mu} \sim \SI{0.1}{\pascal^2}$.
	}
	\item{The correlation length $\xi$ of the fluctuations in the material constant was determined by fitting the covariance of two continuum dipoles of a given separation (calculated using Eq. (\ref{eq:cmn})) to the covariance measured for two network dipoles in Ref. \cite{Beroz2017}. This fit yielded a value $\xi \sim \SI{5}{\micrometer}$. }	
	
\end{itemize}

\section{Probing a Winkler foundation with a finite correlation length} \label{sec:winklercor}

In the main text, we considered the limit $\xi \ll d$ for simplicity. However, our theoretical framework can also describe media with correlation lengths $\xi$ comparable in size to the sensor. In this section, we revisit sensing for the Winkler foundation without making the assumption that $\xi$ is vanishingly small compared to the sensor radius $a$.

By analogy to Sec. \ref{sec:probewinkler} of the main text, the sensor again obtains an unbiased estimate of $\lambda_0$ using Eq. (\ref{eq:haha}). In this case, however, the variance $\delta \lambda_0^2$ is given by:

\begin{equation} \label{eq:deltacorwink}
\delta \lambda_0^2 = \frac{\int \int \langle \delta \lambda(\boldsymbol{r}_1) \delta \lambda(\boldsymbol{r}_2) \rangle \psi(\boldsymbol{r}_1) \psi(\boldsymbol{r}_2) d\boldsymbol{r}_1 d\boldsymbol{r}_2 }{\left(\int \psi(\boldsymbol{r}) d\boldsymbol{r} \right)^2}.
\end{equation}

\noindent To compare to our results from Sec. \ref{sec:probewinkler} of the main text, we take $\psi(\boldsymbol{r})$ to be uniform, as in Eq. (\ref{eq:psiint}). For this choice of probe intensity, we can analytically compute the integrals in the above equation for correlations of the following form:

\begin{equation} \label{eq:fincor}
\langle \delta \lambda(\boldsymbol{r}_1) \delta \lambda(\boldsymbol{r}_2) \rangle  = \Delta_{\lambda} e^{-|\boldsymbol{r}_1 - \boldsymbol{r}_2| / \xi}.
\end{equation}

\noindent Inserting Eqs. (\ref{eq:psiint}) and (\ref{eq:fincor}) into Eq. (\ref{eq:deltacorwink}) gives:

\begin{equation}
\delta \lambda_0^2 = \Delta_{\lambda}V^{-2} \int \int  e^{-|\boldsymbol{r}_1 - \boldsymbol{r}_2| / \xi} d\boldsymbol{r}_1 d\boldsymbol{r}_2 .
\end{equation}

\noindent In three dimensions, we compute these integrals by switching to spherical coordinates to find:

\begin{equation}
\delta \lambda_0^2 = \frac{3\Delta_{\lambda}\xi^3}{2a^6} \left( 4a^3 - 9a^2 \xi + 15 \xi^3  - 3 e^{-2a\xi} (a+\xi)(2a^2 +  5a\xi + 5\xi^2)\right).
\end{equation}

\section{A lower bound on the variance $\delta \lambda_0^2$} \label{sec:boundvar}

In this section, we prove that separately minimizing the configuration of $\psi(\boldsymbol{r})$ in the interior and the exterior can yield a lower bound $\delta\lambda_{0,\mathrm{low}}^2$ on the true minimum $\delta\lambda_{0,\mathrm{min}}^2$ of the variance $\delta \lambda_0^2$. The minimum variance $\delta\lambda_{0,\mathrm{min}}^2$ is given by:

\begin{equation}
\begin{aligned}
\delta\lambda_{0,\mathrm{min}}^2 =\ & \underset{V_f, V_w}{\text{minimize}}
& & S(V_f, V_w) \\
& \text{subject to}
& & \mathcal{C}_1(V_f, V_w, V_f, V_w),\ \mathcal{C}_2(V_f, V_w),
\end{aligned}
\end{equation}

\noindent where $S = \int_{\mathcal{R}} (\nabla V_f \cdot \nabla V_w)^2 d\boldsymbol{r}$ is the unconstrained action for $\delta\lambda_0^2$ integrated over all of space $\mathcal{R}$, the constraint $\mathcal{C}_1(V_f^{(i)}, V_w^{(i)}, V_f^{(j)}, V_w^{(j)})$ is a function of two configurations $i$ and $j$ of the probe potentials that fixes the normalization of the probe intensity:

\begin{equation}
\int_{\mathcal{R}_{\mathrm{int}}} (\nabla V_f^{(i)} \cdot \nabla V_w^{(i)}) d\boldsymbol{r} + \int_{\mathcal{R}_{\mathrm{ext}}} (\nabla V_f^{(j)} \cdot \nabla V_w^{(j)}) d\boldsymbol{r} = 1,
\end{equation}

\noindent and the constraint $\mathcal{C}_2(V_f, V_w)$ enforces the constraints imposed by the finite size of the probe, i.e. that:

\begin{equation}
\nabla^2 V_f = 0,
\end{equation}

\begin{equation}
\nabla^2 V_w = 0,
\end{equation}

\noindent for $\boldsymbol{r} \in \mathcal{R}_{\mathrm{ext}}$. This minimization procedure yields the true, optimal probe potentials $V_f^{(A)}$ and $V_w^{(A)}$. Thus, $\delta\lambda_{0,\mathrm{min}}^2$ is given by:

\begin{equation} \label{eq:lammin}
\delta\lambda_{0,\mathrm{min}}^2 = S(V_f^{(A)}, V_w^{(A)}).
\end{equation}

\noindent To determine a lower bound on this quantity, we start by separating the variance into the following sum:

\begin{equation} \label{eq:decomposition}
S = S_{\mathrm{int}} + S_{\mathrm{ext}},
\end{equation}

\noindent where the contributions $S_{\mathrm{int}}$ and $S_{\mathrm{ext}}$ are given by:

\begin{equation}
S_{\mathrm{int}} = \int_{\mathcal{R}_{\mathrm{int}}} (\nabla V_f \cdot \nabla V_w)^2 d\boldsymbol{r},
\end{equation}

\noindent and

\begin{equation}
S_{\mathrm{ext}} = \int_{\mathcal{R}_{\mathrm{ext}}} (\nabla V_f \cdot \nabla V_w)^2 d\boldsymbol{r}.
\end{equation}

\noindent The action $S$ can be separated in this manner because we have assumed that the probe potentials $V_f$ and $V_w$ are continuous. This continuity ensures that the probe intensity $\psi(\boldsymbol{r})$ cannot diverge anywhere in space, and thereby precludes any additional contributions to the right hand side of Eq. (\ref{eq:decomposition}) from the boundary $\mathcal{B}$. In what follows, we will show that a lower bound $\delta\lambda_{0,\mathrm{low}}^2$ on the variance is obtained by separately minimizing the probe potentials in the interior and exterior as follows:

\begin{equation} \label{eq:dlamlow}
\begin{aligned}
\delta\lambda_{0,\mathrm{low}}^2 =\ & \underset{V_f^{(i)}, V_w^{(i)}, V_f^{(j)}, V_w^{(j)}}{\text{minimize}}
& & S_{\mathrm{int}}(V_f^{(i)}, V_w^{(i)}) + S_{\mathrm{ext}}(V_f^{(j)}, V_w^{(j)}) \\
& \text{subject to}
& & \mathcal{C}_1(V_f^{(i)}, V_w^{(i)}, V_f^{(j)}, V_w^{(j)}),\ \mathcal{C}_2(V_f^{(j)}, V_w^{(j)}).
\end{aligned}
\end{equation}

\noindent This minimization procedure yields probe potentials in the interior ($V_f^{(B)}$ and $V_w^{(B)}$) and in the exterior ($V_f^{(C)}$ and $V_w^{(C)}$). Thus, $\delta\lambda_{0,\mathrm{low}}^2$ is given by:

\begin{equation}
\delta\lambda_{0,\mathrm{low}}^2 = S_{\mathrm{int}}(V_f^{(B)}, V_w^{(B)}) + S_{\mathrm{ext}}(V_f^{(C)}, V_w^{(C)}).
\end{equation}

\noindent Clearly, the following inequality must hold:

\begin{equation} \label{eq:myineq}
S_{\mathrm{int}}(V_f^{(B)}, V_w^{(B)}) + S_{\mathrm{ext}}(V_f^{(C)}, V_w^{(C)}) \le S_{\mathrm{int}}(V_f^{(A)}, V_w^{(A)}) + S_{\mathrm{ext}}(V_f^{(A)}, V_w^{(A)}),
\end{equation}

\noindent because taking the probe potentials in Eq. (\ref{eq:dlamlow}) to be $V_f^{(i)} \rightarrow V_f^{(A)}$, $V_w^{(i)} \rightarrow V_w^{(A)}$, $V_f^{(j)} \rightarrow V_f^{(A)}$, and $V_w^{(j)} \rightarrow V_w^{(A)}$ satisfies the constraints and thereby provides a candidate solution for $\delta\lambda_{0,\mathrm{low}}^2$. Accordingly, probe potentials that do not satisfy Eq. (\ref{eq:myineq}) can only increase $\delta\lambda_{0,\mathrm{low}}^2$, and so would not satisfy Eq. (\ref{eq:dlamlow}). Thus, the above inequality, taken together with Eq. (\ref{eq:lammin}), implies:

\begin{equation}
\delta\lambda_{0,\mathrm{low}}^2 \le \delta\lambda_{0,\mathrm{min}}^2.
\end{equation}

%\bibliography{true-limits}

\end{document}